\documentclass[letterpaper,fleqn,twoside,12pt]{report}
\usepackage[letterpaper, top=1.0in,bottom=1.0in,left=1.0in,right=1.0in]{geometry}

\usepackage{graphicx}
\usepackage{IEEEtrantools,amsmath,amsfonts,amssymb}
\usepackage{mathtools}
\usepackage[titletoc]{appendix}
\usepackage{textcomp}

\usepackage{cite}
\usepackage{hyperref}
\usepackage[nottoc]{tocbibind} 
\usepackage{notoccite}

\usepackage{tocvsec2} 
\DeclareMathOperator{\erfc}{erfc}
\DeclareMathOperator{\erf}{erf}

\usepackage{booktabs}
\usepackage{pdflscape}
\usepackage{multirow}
\usepackage{cellspace}
\usepackage[flushleft]{threeparttable}
\usepackage[table]{xcolor}
\usepackage[strict]{changepage}

\usepackage{graphicx}
\graphicspath{ {images/} }
\usepackage{float}
\usepackage{caption}
\usepackage{subcaption}
\usepackage[export]{adjustbox}

\begin{document}
\bibliographystyle{IEEEtran}


\thispagestyle{empty}
\begin{center}
\begin{minipage}{0.75\linewidth}
	\centering
	\vspace{3cm}
	{\uppercase{\Large Math-aware search engines: \break Physics applications and overview}\par}
	\vspace{3cm}
	{\Large Deanna Pineau\par}
	\vspace{3cm}
	{\Large Aug 2016\par}
	\vspace{3cm}
	University of Waterloo\\
	Waterloo, ON\\
	\vspace{3cm}
	Mini-Thesis / Research Paper\\
	In partial fulfillment of:\\
	MSc Physics\\	
\end{minipage}
\end{center}
\clearpage


\chapter*{Acknowledgments}

This work is dedicated to my parents and to my partner in life.

I would also like to thank my friends and other family members for their support.  For their effort (including reading this document), guidance and patience, I would like to thank my advisor, Brian McNamara (University of Waterloo), and my second reader, Erik Schnetter (Perimeter Institute).  To all the students and professors whom I have met and learned with: thank you for knowledge, discussions, wisdom, and fun times.  I would also like to thank the administrative staff for their efforts in making the clockwork tick.

\chapter*{Abstract}

New search engines exist which take equations as their query, and return results which match either the query's mathematical meaning or its structural presentation.  These results are drawn from scientific papers, online encyclopedias, and math discussion forums.  Their scope includes some of physics, a lot of math, and some content from other natural sciences.

Development of math-aware search engines has progressed far in the decade or so that the field has been alive. The engines are able to parse some equivalencies of mathematics - such as commutativity - and to find similar (in addition to equivalent) formulae to the query.  Their math interpretation abilities, excepting one encyclopedia-based engine (Wolfram Alpha), do not yet reach that of Computer Algebra Systems (CAS) like \emph{Mathematica}.  Developments are proceeding for journal-based engines to include the mathematical meaning-comparison features of CASs; hopefully, in time, the math rules encompassed by CASs will grow past their present limitations in this application as well.

Physics uses math to encode systems' evolutions.  Sometimes, mathematical roadblocks are encountered in physics: a math problem is encountered for which no-one who knows about the problem also knows of a solution.  Searching across disciplines, using equations, might help the researchers to locate a solution.  The fact that jargon is avoided in the search phase allows this cross-disciplinary research to be efficient.  Pre-made solutions are not the only thing which might be found: good approximations, worked-out work-arounds to using that equation, and analogous systems from which to improve understanding (or design experiments) might be found.

To bridge the gap between developers' publications on their search engines, and potential users' awareness of those same engines, this paper reviews their applicability to physics and presents the current math-aware search engines.

\tableofcontents
\listoftables

\chapter{Introduction}
\label{chap:introduction}

What do the AdS/CFT Correspondence, dumb holes, and effective theories have in common?

All provide opportunities for better understanding and/or enable solutions to be found to troublesome math.  The first two do that via linking together pairs of previously disparate phenomena, enabling calculation for quantum field theory from the former, experimentation upon black hole analogue systems from the latter, and insight into physics from both.  \cite{Maldacena-AdS-prop, AdS-CFT-Hubeny, Unruh-Dumb-Holes, Experiment-Dumb-Holes, Carroll-Dumb-Holes}  The latter - effective theories - builds applications and understanding through finding mathematical models which work for a certain setup, yet which do not have to connect with (and can, in fact, blatantly ignore certain parts of) all other known physics \cite{Wells-effective-theories-in-physics-book, Seiler-Stamatescu}.

There are new tools available which enable the finding of more discoveries such as these ones.  They are math-aware search engines.  Like keyword-based search engines, they let the user specify equations - and keywords - to be found in a database of documents.  This gives researchers a new window into interdisciplinary search, by removing jargon and allowing direct searches in our highly-communicative language of math.

So far, their databases include journal articles, encyclopedias, and math forums. \cite{MWS-05-2012, zbMath-database-about, EuDML-web-base, Springer-LatexSearch-post, DLMF-website, DLMF-AMS-notice, Weisstein-Alpha, Wolfram-Alpha-web, Tangent-pub-2015, uniquationWeb}. 

\section{What is math-aware search?}

The goal of these math-aware search engines is to retrieve mathematical formulae which are relevant to a queried formula. \cite{Guidi-Cohen-2015, ZY_math-sim-search, Kohlhase-Sucan-early-MWS, Tangent-technical-v0.3}

What is relevant is a subjective matter. \cite{Guidi-Cohen-2015}  There are two camps at present: one which ranks equations' relevance based on their structural similarities to the query equation(s) (called ``presentation-based''), and another which retrieves equations which have the same mathematical meaning as does the query despite differing notation (called ``meaning-based'').  Each find not only equations which are exactly matching the query, but also those which share similarities with it. \cite{ZY_math-sim-search} For example, a retrieved result might contain only part of the query equation, or might append terms. \cite{Guidi-Cohen-2015, Liska-PhD-Prop, DLMF-AMS-notice}

The search engines are still developing, having been made primarily in the past decade or so. \cite{Guidi-Cohen-2015}  They are well-enough developed to be useful, by providing a new window into our existing research.  As an encyclopedia editor said: ``we don't even know how much we don't know we know''. \cite{Hazewinkel-MKM-is-needed} These search engines help us find what we already know, by enabling us to search in one of our main languages: that of math.

\section{Ability to recognize and compare math}

Presentation-based search might match, for instance, $\log(\sqrt{z})$ and $\log(b)$.  They allow variables' names to be changed, constants and variables to be exchanged, expressions to be substituted for variables, and algebraic terms to be added or removed. \cite{Guidi-Cohen-2015, DLMF-AMS-notice, Pres-vs-Content-2014, Liska-Th ,Liska-PhD-Prop}

Meaning-based search, on the other hand, could ideally match $\sin(x)$ to $(1/2) i (e^{-iz}-e^{iz})$. \cite{Liska-PhD-Prop, Oviedo-Th, Wolfram-Alpha-web}  However, at present, search engines' knowledge of such math equivalencies vary \cite{Guidi-Cohen-2015} and all fall short of being able to match formulas based just on their meaning, ignoring their notation. \cite{Liska-PhD-Prop, MWS-05-2012} A clear delimitation of what they can recognize would require a full understanding of their mathematics parsing algorithms, and is beyond the scope of this paper.  However, a few limitations and examples are explicitly stated in the search engines' publications, and are included in this paper (Chapter \ref{chap:extent-abilities}).  One simple example of what they can parse is commutativity: $ab$ equals $ba$.  Beyond that simple relation, the two web-available meaning-based search engines vary in their abilities.  One can parse all mathematical equivalencies which \emph{Mathematica} can (Wolfram Alpha) \cite{Weisstein-Alpha}, which is still limited when trying to recognize special functions (such as that $T_n(z) = \frac{\delta_{n,0}}{2} + \frac{n}{2}\sum^{\lfloor (n/2) \rfloor}_{k=1} \frac{(-1)^k (n-k-1)! (2 z)^{n-2k}}{k! (n-2 k)!} + 2^{n-1} z^n /; n \in \mathbb{N}$ \cite{Cheb-I-WF} and $T_n(x) = \cos(n \cos^{-1}(x))$ \cite{Cheb-I-WA} are the same function).  Another (MathWebSearch) allows associativity, lets variables' names vary using wildcards, and generalizes some functions to $f(x)$ notation. \cite{MWS-thesis-hasegan, IancuEtAl2014, MWS-05-2012}  However, this engine does not parse all math to a standardized, mathematically-equivalent form: a search for $\lambda x (1 - x + 1 - 1)$ retrieves different results, for example, than does a search for $\lambda x (1 - x)$.  Math-aware search developers are investigating improving their mathematical comparison abilities with Computer Algebra Systems (such as \emph{Mathematica}): the CAS could be used to simplify all database and queried expressions to a standard notation, using rules of math and definitions of functions, so that equations with different writings but the same meaning could be much better found.  \cite{Oviedo-Th, Liska-PhD-Prop}

\section{Database coverage}

The databases covered include physics and math papers \cite{zbMath-database-about, EuDML-web-base, Springer-LatexSearch-post}, special function compendiums \cite{DLMF-website, DLMF-AMS-notice}, encyclopedias \cite{Weisstein-Alpha, Wolfram-Alpha-web}, and math forums \cite{uniquationWeb}.  Not all physics publications are included, yet, but one math-aware search architecture is stated to be ready for full deployment on the arXiv.org content. \cite{MathWebSearch-info}  Some of Springer's articles are covered \cite{Springer-LatexSearch-post, Springer-main-page}, and many European math and mathematical physics journals are math-searchable \cite{EuDML-web-base}.

\section{Applications to Physics}

It has been said, prior to the start of engines' development (roughly a decade ago), that it is easier to re-derive a medium-difficulty result in math than it is to find it in the literature. \cite{Hazewinkel-MKM-is-needed} The author commented that the same could probably be said for applied mathematics disciplines, of which physics is one.  There is much lying around in the literature which each researcher knows little or nothing about.  To find out what we already know, find connections between systems, discover solutions, and target searches and collaborations, the math-aware search engines fill a long-empty niche.  Interdisciplinary and intra-field information finding becomes simpler, because researchers don't have to know the field's jargon to find information: we can search using formulae, instead, to get a starting point.

\section{Paper layout}

Applications to the research of physics, from these search engines, are numerous - some proposed ones are included in Chapter \ref{chap:physics-applications}.  Readers who just want to use the engines can skip Chapter \ref{chap:extent-abilities}, which details the engines' math comparison abilities, limitations, and future developments.  Successful searches and choice of engine are among the subjects overviewed in Chapter \ref{chap:guide-to-use}, which is a quick ``Guide to Use''.  Following that is a usage-oriented summary of each math-aware search engine which is presently web-hosted, in Chapter \ref{chap:avail-SEs}.  That chapter also summarizes the web-available engines' downloadable versions (when applicable), from which custom databases can be created.  Furthermore, it presents a CERN project which is underway, to create math-aware search which is catered to physics' particular needs.  In chapter \ref{chap:screenshots}, screenshots of searches with each search engine are included.

Good reference tables, for which math-aware search engines have each feature, how to query with them, and where to find them, are tables \ref{table:Web-available-main} and \ref{table:web-available-query}.

Finally, a glossary is located in Appendix \ref{app:glossary}, followed by a partial list of other useful math tools which this author encountered during this research (Appendix \ref{app:math-tools}).

\chapter{Physics Applications}
\label{chap:physics-applications}


Equations encapsulate a crucial part of the knowledge in physics \footnote{Although some researchers think we could eventually do physics without using any math, as discussed in submissions (including winning entries) for the 2015 FQXi essay contest, ``Trick or Truth: the Mysterious Connection Between Physics and Mathematics''. \cite{fqxi-essays-2015}}.  When applied to physical systems, they enable prediction and description of phenomena.

Finding information by searching for an equation based on its mathematical meaning is powerful.  It can enable new solutions to physics problems by exposing math solution methods, alternative mathematical models, and work-arounds to using a certain equation.  It can deepen our understanding, and provide new information for development of theories, by allowing researchers to more easily find connections or patterns, analogue systems, and alternative perspectives about a certain model or system.

Additionally, finding information based on equations' structural appearance enables users to look up equations of a specific form - which can carry special significance to a certain field - and to find new connections, solutions, and similar models or systems to those which the researcher is familiar with.

These math-aware search systems are an aid to collaboration, as well as a method of individual search across and within disciplines.  They provide a way for each researcher to better locate information within our known mathematics.  With their aid, we can enhance our interaction with the equations of our observations, models, and established theories.

This section is divided into those applications which are possible now, and those which require some significant developments.  Most of the applications herein are the ideas of this author, except when noted otherwise, although it is questionable whether some of them are \emph{implied} in other publications which are cited elsewhere in this work.  Ideas included here are intended to give physicists a start of an idea on how these math-aware search engines can be used in everyday research.

\section{Now Available (Although they improve with time)}

\subsection{Solve Math Problems}
Many times in physics, we encounter equations which impede progress.  The researchers who work on them can not find the solutions to them.  These equations might have solutions (e.g. approximations, work-arounds, or more useful - mathematically equivalent - ways of expressing them) which are known to certain fields of math, or which have been used elsewhere in physics or the other sciences.  Yet, finding them has so far been prevented by the use of jargon and the inefficiency of identifying collaborators who know something about the solution.  Math-aware search engines help a researcher to identify systems, fields, and collaborators which would enable the system of interest to be better understood and/or applied.  Good examples of this inter-disciplinary approach benefiting physics include the AdS/CFT correspondence and holographic duality theories.  To expand upon this point, it has been said (in math) that ``we don't even know how much we know that we don't know we know'' \cite{Hazewinkel-MKM-is-needed}: finding what is applicable out of what we know is a good approach to make progress.

The development of emergent theories, fundamental physics, effective theories, applied physics, experiments, and phenomenology can all benefit from such improvements.

\subsection{Design Models and Match them with Physical Systems}
Finding mathematical models which are useful for a case can be done by first coming up with a model which is reasonable but hard to solve, then researching ways to make it more tractable.  The first model which one comes up with need not, then, be the best model: searching the literature can locate refinements which make it more solvable.  This might include avoiding the equation in question by changing the modeling setup, rewriting the equation into an equivalent but manageable format, or making approximations which other researchers have found to work well.  Context will need to be considered, of course, but having this start at discovering better models is priceless.  Keyword-based search engines can help with this, but they are inherently limited by varying terminologies between fields (i.e. jargon) and the need to know what kind of systems one wants to find.  These equation-based search engines enable researchers to find mathematical models using the most appropriate language: that of math.  They already know some of the ``synonyms'' of math (such as $a+b = b+a$), and will only improve with time.

\subsection{Find Analogous Systems for Experimentation}
Other times, we find it useful to experiment with analogous systems to ones which we can not directly observe.  For instance, an investigation occurred into Black Holes and their particle production via an analogous system: dumb holes. \cite{Unruh-Dumb-Holes}  These sound-based analogues of black holes were experimented upon to better understand and predict black holes' behaviours. \cite{Experiment-Dumb-Holes, Carroll-Dumb-Holes}  Another set of experiments has been pursued in order to explore additional fundamental predictions of quantum field theory, such as the Sauter-Schwinger effect; \cite{Schuetzhold-fund-experiments} discusses them in a colloquium.

\subsection{Pattern-find within physics}
Another application of finding analogous systems is to explore what physical theories might connect systems which are found to be analogous.  For instance, one might find patterns for merging gravitation with quantum physics, by searching physics publications for quantum or gravity equations.  While this application is limited, so far (because the math-aware search engines are not yet advanced enough to identify mathematically-equivalent notation), it still provides an excellent start.

\subsection{Find a specific format of an equation}
The form in which an equation is written can have significance.  For example, writing the Schrodinger equation as a probability current indicates that it might be under study as a fluid flow. While this particular format has a name (``probability current''), not all significant formats of equations have a name.  These un-named equations can be difficult to look up - as can some of the named ones - without knowing the jargon of the field.  Math-language search engines let researchers get around the jargon, during the search phase.

\subsection{Identify Special Functions}
If a researcher can identify that a function of interest is a special function, then new resources become available with which to solve it.

However, this application is not well-developed yet.  Although databases of special functions exist, their math-aware search engines do not recognize all possible ways of writing special functions.  For instance, the definitions of the special function ``Chebyshev polynomial of the first kind'' are different on each of Wolfram Alpha, DLMF, and Wolfram Functions (from which the ``primary definition'' was taken); none of the special function databases' search engines (i.e. Wolfram Alpha and DLMF) recognize the definitions which came from one of the other three websites.  Yet, it can work to identify special functions via this route, given present capabilities: both Wolfram Alpha and the DLMF recognized the complementary error function's ($\erfc(x)$) definitions which were drawn from the same three websites.  These experiments with special functions are described in slightly more detail in section \ref{sec:math-meaning-limitations}.

For now, using a search engine of this type to identify special functions is limited: it can provide a positive identification, but can not tell a researcher that a given function is \emph{not} a special function.  As math-aware search engines get better at translating between different math notations, this application will be more complete.

\subsection{Look up partially-remembered theorems/laws}
Long-ago learned equations might be remembered for their application, but the formula and name can be difficult to fully bring to mind.  If enough of an formula is remembered, it can be looked up in a math-aware search engine to (hopefully) find the full version of that formula. \cite{MWS-05-2012}

As an example (adapted from \cite{MWS-05-2012}), suppose that a researcher remembers an integral inequality exists for expressions of the form $\int_D |f(x) g(x)| dx$.  However, the researcher does not remember its name or full equation.  The inequality which they half-remember is Holder's inequality:
\begin{equation}
	\int_D |f(x) g(x)| dx \le \Bigg( \int_D |f(x)|^p dx \Bigg)^\frac{1}{p} \Bigg(\int_D |g(x)|^q dx \Bigg)^\frac{1}{q}
	\label{eq:holders-ineq}
\end{equation}
The researcher might try a keyword search for ``integral inequality absolute value'', or ``Integral $|$f(x)g(x)$|$ approximate'', or similar phrases.  However, for these search terms the first page of Google returns nothing which is obviously related to Holder's inequality.  Using math-aware search engines can return much more relevant results.  For instance, the search engines included herein which are powered by MathWebSearch \cite{MWS-05-2012} - that is, zbMATH \cite{zbMATH-engine} or MathWebSearch's online demo \cite{Sentido-search} - can locate Holder's inequality given the input $\int_D |f(x) g(x)| dx$.

\subsection{Look up the meaning of unfamiliar math notation}
Especially when doing interdisciplinary research, it is easy for a researcher to come across notation which he or she is unfamiliar with.  The context is not always explained adequately, in the text, for that researcher to easily find the notation's meaning: field's conventions and ``common knowledge'' are often not stated in publications. \cite{DavenportEtAl2003, Rec-and-retrieval-rev}  Math-aware search engines which use equations' structure (i.e. ``Presentation-based'') to search for results can be used to find more information about the unfamiliar notation.  For example, the search engine ``Tangent'' proposes this application as a use of its software for both students and researchers \cite{Tangent-thesis}; for students, they give the example that the binomial coefficient's notation ($^4_2$) could be drawn in $m_{in}$ (section \ref{sec:min-description}) and subsequently searched for \cite{Tangent-pub-2015,Wangari-et-al_2014_paper}.  For researchers, an application might be to look up why $\pi(2) = 1$, which applies when discussing the ``prime counting function''. \cite{Wangari-et-al_2014_paper}

\subsection{Preparing LaTeX Equations for Publication}
For complex or highly-formatted LaTeX equations, it can be useful to either copy and paste an existing equation from another researcher's publication, or generate it via a drawing or template editor.

Copy and paste is now possible for a wider range of publications.  It has been possible on some arXiv.org publications for some time: users can download the source code for a publication, which includes the LaTeX file \cite{arxiv-unpack}; this is somewhat tedious, but can be worth it.  A simpler method applies, now, to some of the journal Springer's publications. Using Springer's LaTeXSearch website, researchers can bring up all equations in a paper, by searching for its DOI or title.  Results will include an image and LaTeX code for each equation.  Finding LaTeX code in this way, for direct copying, is one of the main purposes which drove LaTeXSearch to be made \cite{LatexSearchAbout}.

To generate LaTeX code for an equation, one can also draw it on the computer (such as in $m_{in}$ \cite{min-interface}) or use a template editor such as those described in section \ref{sec:template-editors}.

\section{Future Possibilities}

\subsection{Discover Alternate Interpretations via Different Equation Formats}
Rewriting an equation can change the way a person understands it.  For example the Schrodinger wave equation can seem less (or perhaps more) mysterious when it is written as a fluid flow in probability:
\begin{equation}
	-\frac{\hbar}{2 i m} \nabla \cdot (\Psi^* \nabla \Psi - \Psi \nabla \Psi^*) = \frac{d}{dt}(\Psi^* \Psi)
	\label{eq:Schro-fluid-second-writing}
\end{equation}
(This particular form of the Schrodinger equation is one whose structure arises in any theory within which a conservation law applies to an extensive quantity (such as mass, charge, or heat energy).  However, let us pretend for a moment that we don't know that connection, yet, and are looking for ways in which certain branches of physics relate to one another - such as in quantum gravity.)

A math-aware search engine could find connections like this, wherein a standard equation is rewritten to a format which matches a different branch of physics (here, fluid mechanics).  This discovery of how an equation fits into the current system of physics can seed new depth to understanding.

As math-aware search engines are able to recognize more of math's equivalencies, this application becomes better.  However, a researcher using the present-day math-aware search engines would have a very difficult time finding these connections using them.

\subsection{Discover new solution methods via Different Equation Formats}
Writing an equation in a new way can also elucidate paths towards solving it.  A simple example of this in everyday research is the use of substitutions in integrals, or integration by parts.  Finding other researchers' writings of an equation gives the researcher an opportunity to see what other forms of writing other researchers might have found useful: it gives them direction towards finding a potential solution.

However, this application is only possible when the mathematical meaning of formulas can be recognized by math-aware search engines, to an extent which lets them match equivalent formulas which are written in different ways.  Their present level of meaning-extraction is probably far too low to enable this use, although their in-development upgrade to parse meaning using \emph{Mathematica}-type software will likely make this feasible. 

Improvements to this application are possible with certain feature developments in the math-aware search engine.  For instance, if the results could be \emph{ranked according to solvability} - that is, whether an equation can and/or has been solved analytically (or numerically, in an efficient manner) - then researchers could find ways of writing their equation such that solutions fall out easily.  This goes beyond \emph{Mathematica}, since even \emph{Mathematica} (and similar software) sometimes needs equations to be written in different representations in order to solve them.  Furthermore, \emph{Mathematica} can not change how the mathematical model is attached to the physical world (e.g. is $x$ the linear distance from a point, or the distance along a curved wire?), which a researcher who views equation results could do.  When a researcher searches existing models, the searcher can also see what assumptions were useful to those models, and decide whether they can apply to the system at hand; \emph{Mathematica} requires many such assumptions to be included as input to the equation.  Finding information in the literature can therefore be of benefit beyond what \emph{Mathematica}-like systems provide, for solving equations.  If these search engines could determine whether a given equation format is solvable, or read that data from the document which includes it, then finding solvable mathematical models (or just equations) in the literature becomes easier.

\subsection{Speed up computer programs' run-time}
Some analytic equations run faster than others, either due to the number of steps required to compute them or to the run-time engine's specific architecture.  As a simple example, rewriting
\begin{multline}
	f(x) = \cos(a)\cos(b)\cos(c) \\
	-\sin(a)\sin(b)\cos(c) -\sin(a)\cos(b)\sin(c)-\cos(a)\sin(b)\sin(c)
\end{multline}
to
\begin{equation}
	f(x) = \cos(a+b+c)
\end{equation}
will cause less computations to be performed when the code runs.  Although the gain is small when solving this equation once, if it must be solved millions or billions of time during run-time then the small changes will add up.

Math-aware search engines can help to find \emph{useful} re-writings of an equation, which result in faster run-time: there are many ways to rewrite equations, but the format which another researcher used may have been chosen because that researcher found it to work well for the application.  Other researchers might have already put in the work to sort through some of the less efficient writings.  Equivalent mathematical notation might be found in this way, in addition to approximations (which lose precision when used).

Using math-aware search engines for this task works best if the math-aware search engine ``knows'' a lot of mathematical equivalencies.  Then, notations of the equation which are not obvious from its input format could be found.  It is possible to do this application with current math-aware search engines, but it would be very limited due to their meaning-based comparisons limitations.

\subsection{Custom ``alerts'' when new publications contain a specific equation}
Routing of results - that is, updating users when new documents are available which includes their equation query - is one proposed use of math-aware search engines, by \cite{Youssef-Roles}.  When journal articles comprise the database, this would mean that researchers can set up ``alerts'' to be sent to them when a new article with a certain equation is published (and is indexed into the search system's database).

\chapter{Extent of Engines' Abilities}
\label{chap:extent-abilities}


Building a math-aware search engine is no easy task, but it has been accomplished to some degree.  Over the past thirteen years \cite{Guidi-Cohen-2015}, these search engines have gained the ability to interpret some basic math, to use equations' structure and appearance to find results, and to retrieve both exact and simply similar matches to a queried equation.  They can take keywords alongside multiple equations in a single query, and include (named and un-named) wildcards.  Their database coverage is impressive, and is expected to continue growing.

In the future, their developments are expected to include better extraction of math meaning from equations and their surrounding text \cite{MWS-05-2012,  AizawaEtAl2014}, an improved ability to match equations using their mathematical meaning \cite{Liska-PhD-Prop, Oviedo-Th}, a better understanding of what constitutes ``similar'' equations \cite{ZY_math-sim-search}, improved database coverage \cite{MathWebSearch-info, MWS-05-2012, Tangent-pub-2015, WebMIaS-descr-pub-2014, RakosnikEtAl2014, Kohlhase-Scaling-MWS}, and a focus on user needs and interactions \cite{AizawaEtAl2014}.  Some other neat tools - such as email alerts for new publications which contain a given equation - are also imagined for the future \cite{Youssef-Roles}.

This section discusses the operational steps, current abilities, and future developments for these math-aware search engines.  Beginning with an overview of the engines' generic operation and their database coverage, it then delves into the extent of each engine's abilities to extract and compare the mathematical meaning of equations.  This information complements the usage-oriented presentation of each engine which is found in Chapter \ref{chap:avail-SEs}, by sketching the engines' limitations and rules for retrieving ``similar'' or ``equivalent'' equations.  At the end of this chapter, future developments of this field are discussed, and a few physics-oriented developments are proposed.

\section{Sequence of Events in an Engine's Operation}

Both the query (input) equation and those located in the database's documents are processed in order to match equations in a math-aware way.  This is done for presentation-based and meaning-based search.

Via various algorithms, variables' names and constants' values are either preserved or ignored.  Preserving their values allows exact matches to the query to be found.  Allowing one or both of them to be ignored broadens the search results while preserving the mathematical content of the equation. (although sometimes physical meaning is lost when constants' values are ignored). \cite{SojkaEtAl2011, Oviedo-Th, Tangent-pub-2015}  To account for physical experiments' constants having varying precision, one search engine (in-development) performs iterative rounding on each query's constants. \cite{Oviedo-Th}

Interpreting operators and variable types requires some context about what the equation represents.  For example, the variable $\mathbf{R}$ might represent a matrix, a vector, or the real number line.  A variable might be real, complex, or of another type.  Math-search engines often extract context from an equation's parent document, in order to decipher what the symbols represent. \cite{Oviedo-Th, AizawaEtAl2014}  However, extracting more context is an area of ongoing development. \cite{AizawaEtAl2014, MWS-05-2012}

Equivalence of mathematical expressions must then be determined, between the query expression and those in the database.  Currently, algorithms either base their comparison on the structural similarities between expressions, or the mathematical meaning behind the equations (regardless of their notational representation). \cite{Pres-vs-Content-2014}  The former can be called ``presentation-based'' search (or ``syntactic'' search), and the latter ``meaning-based search'' (or , more properly, ``semantic'' search). \cite{Pres-vs-Content-2014, Liska-PhD-Prop} Meaning-extraction algorithms still contain errors \cite{Pres-vs-Content-2014}, however, and meaning-based comparisons of equations are not yet perfect \cite{AizawaEtAl2014, MWS-05-2012}; developments are underway to improve this \cite{MWS-05-2012, Pres-vs-Content-2014}.  

An additional element which can be considered is the complexity of the equation, so that equations with similar complexity to the query are put nearer to the top of the results. \cite{SojkaEtAl2011}

\section{Database coverage}

Academic papers (such as journal articles), a curated encyclopedia, Wikipedia, and mathematics forums are all accessible via math-aware search engines. \cite{Springer-LatexSearch-post, EuDML-web-base, zbMath-database-about, Weisstein-Alpha, dprl-descriptions, uniquationWeb}

The papers include many from physics, a large number from math, and some from other applied mathematics sciences.  Their sources include the journal Springer (in LaTeXSearch), and the numerous journals which are accessed by the search services EuDML and zbMATH. \cite{Springer-LatexSearch-post, EuDML-web-base, zbMath-database-about}  A portion of the arXiv is searchable by equation, but it is only for demonstration purposes; however, a developer has stated his team's engine (MathWebSearch) to be ready to handle the entire arXiv. \cite{MathWebSearch-info}

Encyclopedic and discussion-based sources cover some of physics, and a good portion of math.  A curated encyclopedia is maintained by Wolfram Alpha, which brings up information about some physics topics when their appropriate equation is searched for. \cite{Weisstein-Alpha, Wolfram-Alpha-web}  Wikipedia's mathematical pages have been partially indexed by Tangent, although their content does not update when Wikipedia does \cite{dprl-descriptions}; hopefully this coverage extends soon.  The math forums covered include planetmath.org, mathoverflow.net, and \linebreak math.stackexchange.com, with searching provided by Uniquation. \cite{uniquationWeb}

Scalability is still a challenge which many of the math-aware search engines have to overcome. \cite{Tangent-pub-2015,MWS-05-2012}  However, the range of material covered by some of the engines so far is impressive.  In particular, there are over 3 million publications in zbMATH \cite{RakosnikEtAl2014}, and over 250,000 on EuDML \cite{EuDML-web-base}; all of them are searchable using equations.  The over 1.1 million arXiv.org pre-prints are also stated to be serviceable by MathWebSearch, provided that server space is provided for the search engine. \cite{MathWebSearch-info}  The developers of these search engines want to provide math-aware search over additional scientific resources (e.g. \cite{MathWebSearch-info, Kohlhase-Scaling-MWS, RakosnikEtAl2014, WebMIaS-descr-pub-2014}), including servicing databases on the order of current keyword-based search engines' sizes \cite{Tangent-pub-2015}.

\section{Extent of abilities to: extract math in-context}
\label{sec:extent-abilities-extract-math-in-context}

Search engines which act upon journal articles and other web-based sources need to find and extract mathematical equations from their database documents.  Because math is intended to be read by humans, rather than by computers, \cite{DavenportEtAl2003} this requires some algorithmic design \cite{Oviedo-Th, Rec-and-retrieval-rev, AizawaEtAl2014}.

There is a lot of symbol recognition which occurs before mathematical context is considered.  For example, there are many ways to encode equations in LaTeX: a bold symbol could be encoded using ``\textbackslash{}mathbf\{symbol\}'', ``\textbackslash{}pmb\{symbol\}'', or a few other commands. \cite{Latex-symbols-guide}  Individual symbols can also be written in many different ways; the Angstrom symbol in unicode can be written as a ``latin capital letter A with ring above'' (code 0xc5) or ``Angstrom sign'' (code 0x212b). \cite{Oviedo-5e-2pg}  The math-aware search engines account for the different encodings of given symbols, but they do have troubles when those encodings are non-standard (for instance, see the example below concerning ``\textbackslash{}dot'').

To interpret math, these search engines pull context from the text and apply field-specific conventions whenever possible.  For example, the text might specify that $\lambda$ is a binding function, or that it is a variable \cite{Rec-and-retrieval-rev}; or, $f$ might be either a functional operator or a variable (such as frequency).  An integral ($\int f(x) dx$) could be Reimman, Lebesque, or any of the other dozen or so types of anti-derivation. \cite{Kohlhase-Sucan-early-MWS} Vectors, matrices, and other symbols might also be defined in the text, which can resolve ambiguities such as whether $\boldsymbol{R}$ is a matrix, vector, or the real number line.  Extracting meaning from documents is difficult without this context; algorithms for context-extraction are known to improve the relevance of results from math-aware search engines. \cite{AizawaEtAl2014}

Examples of presently problematic notation include subscripts, incorrect notation, and conventions which are specific to geographical regions.  For instance, the expression $x_1^2$ might indicate the square of the variable $x_1$ ($(x_1)^2$), or might be the first entry in a sequence of $x^2$ terms ($(x^2)_1$). \cite{Rec-and-retrieval-rev}  Notation which is encoded in a human-readable but formally incorrect manner is also problematic.  The times operator is a good example of this: it can be properly encoded as ``\textbackslash{}dot'', but can also be written as ``.''.  The latter appears as a decimal point, which is problematic when it appears between numbers.  For example, the the historical formula
\begin{equation}
	\sqrt{2} = 1 + \frac{1}{3} + \frac{1}{3\dot 4} - \frac{1}{3\dot 4\dot 34}
	\label{eq:decimal-problems}
\end{equation}
is here encoded using ``\textbackslash{}dot'', but when the encoding uses ``.'', then the last denominator is interpreted as $3.4.34$ by MIaS.  Aside from this being a senseless number - three point four point thirty-four - it in no way matches the intention of the formula. \cite{Liska-Th}  A good example of a geographic difficulty with conventions is the notation for the binomial coefficient: the Russian convention is $C^k_n$, which is the opposite of the $C_k^n$ French standard. \cite{Kohlhase-Sucan}

To summarize, it is difficult to parse the human-oriented mathematics equations properly.  A lot of progress has been made, and developments continue. \cite{Liska-Th, Pres-vs-Content-2014, NghiemEtAll2013}  Search engines which use context to extract meaning from equations' surrounding text have been noted to retrieve more relevant results. \cite{AizawaEtAl2014}.

\section{Extent of abilities to: match formulae}
\label{sec:math-meaning-limitations}

These search engines take a queried equation, and find equivalent and similar equations to it.  The main algorithms used, in this task, are presentation-based search and meaning-based search.  The applied uses of each are overviewed in section \ref{sec:pres-vs-meaning-usage}.  Within these two categories, the abilities of the available math-aware search engines vary considerably.

In general, most of the meaning-based search engines can understand commutativity ($a+b=b+a$), allow variables' names and constants' values to change, and allow variations in the equations' structure (i.e. adding or dropping terms, allowing entire expressions to take the place of a variable, etc.).  The grand majority of them can not do associativity ($ab = ba$) or distributivity ($a + (b + c) = (a + b) + c$), except by allowing variables' names to vary and terms to be re-arranged.  More complex math equivalencies are generally not included; however, progress is being made to incorporate \emph{Mathematica}-like Computer Algebra Systems into the engines' algorithms, so that all the mathematical rules, simplifications, and other tools of \emph{Mathematica} can be used in identifying which equations are equivalent, mathematically. \cite{Oviedo-Th, Liska-PhD-Prop}

In this section, the abilities of each meaning-based and presentation-based search engine is reviewed.  The meaning-based engines' discussion focuses on the extent to which their algorithms are able to recognize mathematically equivalent expressions which differ visually.  In contrast, the presentation-based engines' discourse considers which criteria the engines define to say which equations are ``structurally similar''.

\subsection{Mathematical meaning-based comparisons}

Meaning-based search algorithms attempt to find equations which math the mathematical meaning of the searched-for (``queried'') equation. \cite{Pres-vs-Content-2014, Liska-PhD-Prop, MWS-05-2012} Each uses a different approach, and all fall short of being able to fully identify mathematical equivalencies given different writings of the math (including Wolfram Alpha).  Nevertheless, they are able to identify mathematical equivalencies to varying degrees.

\subsubsection{Wolfram Alpha}
\label{sec:Wolfram-Alpha-Math-Meaning-Extent}

Wolfram Alpha, which both searches its internal encyclopedia and computes results, uses all the mathematical rules and simplifications which Mathematica knows. \cite{Weisstein-Alpha, Wolfram-Alpha-web}  As such, it is very powerful.  However, it still can not recognize all formats of a special function.  To experiment upon this, the definition of the Chebyshev polynomial of the first kind (hereafter Cheb-I) was found on the Wolfram Functions website to be
\begin{equation}
	T_n(z) = \frac{\delta_{n,0}}{2} + \frac{n}{2}\sum^{\lfloor (n/2) \rfloor}_{k=1} \frac{(-1)^k (n-k-1)! (2 z)^{n-2k}}{k! (n-2 k)!} + 2^{n-1} z^n /; n \in \mathbb{N}
	\label{eq:Cheb-I-Wolfram-Functions}
\end{equation}
(this was the first ``primary definition'' on Wolfram Functions). \cite{Cheb-I-WF} Yet, when this equation was entered into Wolfram Alpha's search query - using a simple copy-paste of the Mathematica text from Wolfram Functions, with a check for errors - Wolfram Alpha failed to parse it.  Various formats of the equation were tried, and eventually Wolfram Alpha recognized the formula - but, it did not identify it as the Cheb-I polynomial.  Wolfram Alpha has its own definition for Cheb-I, which is found by searching by keyword for the function:
\begin{equation}
	T_n(x) = \cos(n \cos^{-1}(x))
	\label{eq:Cheb-I-Wolfram-Alpha}
\end{equation}
\cite{Cheb-I-WA}.  When this equation is entered into the Wolfram Alpha search box, it identifies that it is Cheb-I.  On the other hand, another experiment which was performed with the same method revealed Wolfram Alpha to correctly identify two other websites' definitions of the complementary error function ($\erfc(z)$).  In that experiment, definitions were pulled from the Wolfram Functions website,
\begin{equation}
	\erfc(z) = 1 - \frac{2}{\sqrt{\pi}} \sum_0^{\infty} \frac{(-1)^k z^{2 k + 1}}{k! (2 k + 1)}
	\label{eq:erfc-Wolfram-Functions}
\end{equation}
\cite{Erfc-WF} and the DLMF \footnote{The DLMF was not included for the Cheb-I experiment because it defined Cheb-I in terms of a few other functions, which were defined across many of its webpages, in such a way that no formula for Cheb-I was actually written in its database. \cite{Cheb-I-DLMF}},
\begin{equation}
	\erfc(z) = \frac{2}{\sqrt{\pi}}\int_z^{\infty}e^{-t^2}dt = 1 - \erf(z)
	\label{eq:erfc-DLMF}
\end{equation}
\cite{Erfc-DLMF}.  Wolfram Alpha defined the $\erfc(z)$ to be
\begin{equation}
	\erfc(z) = 1 - \erf(z)
	\label{eq:erfc-Wolfram-Alpha}
\end{equation}
\cite{Erfc-WA}.  When each of the DLMF and Wolfram Functions $\erfc(z)$ definitions were input to Wolfram Alpha, the results identified that the equation was the complementary error function.

\subsubsection{MathWebSearch}
MathWebSearch is the only working math-aware search engine for journal articles which uses meaning-based search algorithms.  Determining the limits of its abilities, as with Wolfram Alpha, would require a full understanding of its algorithms, which is beyond the scope of this paper.  However, many of its features and their limits are explicitly stated in the publications which discuss it.

Variable names can either be preserved (for ``non-query'' variables) or ignored (for \linebreak ``query'', or ``wildcard'' variables). \cite{MWS-thesis-hasegan} This means that a user can enforce commutativity to hold (although it may be encoded into MathWebSearch already - the literature does not mention commutativity) simply by writing their equation using query variables. It supports the use of ``named'' query variables, so that a user can specify both that certain variables should have the same name (as in $ab + a$) and that the actual variable used is irrelevant (so $xy + x$ would also match) \cite{MWS-0.5-fgwm, zbMATH-search-page}. Based upon experimentation with zbMATH \cite{zbMATH-engine}, query variables in MWS are allowed to be replaced with subscripted variables, non-subscript variables, or numbers: $(?x)^2 + (?y)^2$ can retrieve $a^2+b^2$, $x_1^2 + x_2^2$, and $8^1 + 1^2$. Furthermore, when non-query variables are used, the variable names are not allowed to vary at all \footnote{To test whether non-query variables were allowed to vary, the expression ``c\textasciicircum{}2 + d\textasciicircum{}2'' was searched on zbMATH's formula-oriented search page, \cite{zbMATH-engine}. Only two pages of results were found, all of which used exactly the formula $c^2 + d^2$. However, searching for the query variable form of this equation, ``(?c)\textasciicircum{}2 + (?d)\textasciicircum{}2'', brought up 497 results, which included those using variables other than ``c'' and ``d''; examples included $x_1^2 + x_2^2$, $x^2 + y^2$, and $8^2 + 1^2$.}.

Another basic math function, associativity, is also included in MathWebSearch \cite{IancuEtAl2014}.

To better recognize math operators' changeable names, MathWebSearch also generalizes elementary functions such as $\sin(x)$ to generic operators like $f(x)$. \cite{MWS-05-2012} When searching for an equation, it also lets only part of the searched equation match the results, or only part of the result to match the searched-for equation. \cite{MWS-05-2012} These two items combine to make it good for retrieving half-forgotten formulas and finding approximations. For example, consider Holder's inequality:
\begin{equation}
	\int_D |f(x) g(x)| dx \le \Bigg( \int_D |f(x)|^p dx \Bigg)^\frac{1}{p} \Bigg(\int_D |g(x)|^q dx \Bigg)^\frac{1}{q}
	\label{eq:holders-limitations-sec}
\end{equation}
Searching for $\int_D |f(x)g(x)|dx$ could bring up equation \ref{eq:holders-limitations-sec}, if it was in the database. \cite{MWS-05-2012} In addition, a search for $\int_R |sin(x)cos(x)|dx$ could also find equation \ref{eq:holders-limitations-sec}. \cite{MWS-05-2012} However, searching for $\int_R |sin(x) cos(2x)| dx$ will not identify equation \ref{eq:holders-limitations-sec} as a match: the algorithms of MWS interpret the $x$ and $2x$ to be the arguments of the function, causing it to parse this equation's generalization as $\int_D |f(x)g(y)|dx$ and not $\int_D |f(x)h(x)|dx$. \cite{MWS-05-2012}

\subsection{Presentation-based (Structural) Comparisons}

These engines find equations which are structurally similar to the query.  The results retrieved might differ from the query by changed variable names; added, dropped, or re-arranged terms; subexpressions substituted for variables; or other visual-based criteria. \cite{DLMF-AMS-notice, Liska-PhD-Prop, Tangent-technical-v0.3}

\subsubsection{DLMF}
Searching its own database of special functions, the DLMF allows queried equations to bring up results whose structure differs slightly from that of the input equation.  For example, it will add terms which it thinks appropriate: $\sin^2$ will be appended with and $(x)$ to make it $\sin^2(x)$. \cite{DLMF-2007}  If a query returns no exact matches, then the DLMF's engine allows the symbols in the equation to be re-arranged (thus approximating commutativity and associativity) and to be separated by other math terms; this is a form of ``query relaxation''. \cite{DLMF-AMS-notice, Liska-Th}

Unique to DLMF is a ``metadata'' search feature.  It allows a search to contain keywords for certain branches of math (e.g. ``trigonometric''), such that those keywords represent any function from that category (e.g. $\sin(x)$, $\cos(x)$, etc.).  The keyword does not have to appear on the search results page.

For example, a user could search for ``trigonometric\textasciicircum{}2 + trigonometric\textasciicircum{}2'', and retrieve identities such as $\sin^2(x) + \cos^2(x) = 1$.  Due to the DLMF's query relaxation, such a search also returns $\sec^2(z) = 1 + tan^2(z)$, into which terms were inserted between the ``trigonometric\textasciicircum{}2'' terms; $\sin(2z) = (2\tan(z))/(1+\tan^2(z))$, in which one ``trigonometric\textasciicircum{}2'' term was dropped; and many other functions.

\subsubsection{Tangent}
\label{sec:tangent-extent-abilities}
Visual aspects of equations are highly emphasized in Tangent's search over Wikipedia articles. \cite{Tangent-technical-v0.3, tangent2016sigir} For instance, $1 + x^2$ is considered to be different from $x^2 + 1$, although each may come up in the results of the other as one of the ``similar'' equations. \cite{TangentSE}  The engine matches equations based on the relative positions, and on the types, of the symbols therein. \cite{Tangent-technical-v0.3}  Wildcards are allowed in the web-available version (which is not the latest development), but each wildcard can only represent a single symbol. \cite{tangent2016sigir}

Based on experimentation with the online version, equations which are considered ``similar'' to one another can have interchanged variables and constants, differently named variables, constants of differing values, and terms added or removed. \cite{TangentSE} For example, searching for ``g(z) = 0'' returns results which include the following: $g(z) = 0$, $g(E) = 0$, $g(0) = 0$, and $h(z) = 0$.  \cite{Tangent-pub-2015, TangentSE} The latter three of these results all share the same ``score'' - that is, they are considered to have the same degree of similarity to the input equation. \cite{TangentSE}

The web-available demo version of Tangent is an old version of the software (v0.1).  Developers have now released version 0.3.1, which is available for download and use; that version was used for a recent competition in math-aware information retrieval. \cite{dprl-descriptions, ntcir12tangent}  Its upgrades include better discovery of ``similar'' equations, broader wildcard support (including allowing a single wildcard to represent a subexpression, instead of just one symbol), keywords alongside equations in a query, and scalability. \cite{ntcir12tangent, tangent2016sigir, Tangent-technical-v0.3, tangent2014ntcir}  It allows subexpressions to be subsitituted for other subexpressions, regardless of their mathematical meaning or size - for example, $\sin(a + b)$ has the subexpression $(a+b)$, which can be substituted for any other expression inside parenthesis when retrieving results. \cite{ntcir12tangent} Notationally, this updated engine can also query matrices, prefix subscripts (e.g. $_nC$), and prefix superscripts (e.g. $^rC$). \cite{tangent2014ntcir}  Possible upcoming features indicated by the developers include incorporating textual context into equation matching \cite{tangent2016sigir}, improving keyword-vs-equation weighting when ranking results \cite{ntcir12tangent}, and adjusting the weight which is given to matching variables (versus allowing them to vary) when retrieving and ranking results \cite{ntcir12tangent}.

\subsubsection{Math Indexer and Searcher (MIaS)}
\label{sec:MIaS-meaning-abilities}

The developers of MIaS give a fair bit of detail about what their algorithms can do.  Their engine - currently used to search over journal articles - uses presentation-based search but is moving towards incorporating meaning-based algorithms. \cite{Liska-PhD-Prop}

When finding similar equations, the engine allows constants' values and variables' names to change. \cite{Liska-PhD-Prop}  However, it distinguishes between variables with and without subscripts (e.g. $x_1$ vs $x$), and does not allow them to be interchanged.  For example, the query $x_1 + y_1$ would \emph{not} retrieve $x + y$.  \cite{Liska-PhD-Prop}  This treatment of subscripts can be problematic for physics applications, because subscripts are often used to denote which physical characteristics a variable is attached to (such as $x_i$ for the position at a certain time, or $d_s$ for the distance to Saturn); these subscripts have no mathematical significance, except to distinguish one variable from another.  Another equation characteristic which is allowed to vary is that subexpressions in the query are searched for: for example, the query $a + b^{c+2}$ has the subexpressions $b^{c+2}$ and $c+2$. \cite{Liska-PhD-Prop}  Furthermore, the structure of equations is recently able to be considered in matching formulae: $x + y/z$ would structurally match $a + \sqrt{b}/c$, having the form ``item1 + item2 / item3''.  Finally, the mathematical equivalency of commutativity is accounted for, so that $a + 5$ is considered equal to $5 + a$. \cite{Liska-PhD-Prop}  

Some problematic queries, for MIaS, are elucidated in \cite{Liska-Th}.  These were queries which were sent with knowledge of what should be retrieved, based on a human assessment, and which the algorithm failed at.  Some situations where the query did not succeed included: (a) a query used ``$=$'', whereas the target equation used ``$\equiv$'', (b) the query wrote text in math-mode using ``\textbackslash{}text\{\}'', whereas the target did so with ``\textbackslash{}rm\{\}'', (c) the target equation was broken up into multiple lines, whereas the input was on one line. \cite{Liska-Th}

Developments are underway to expand MIaS's ability to identify similar equations and design algorithms which understand a much broader range of math. \cite{mias2016, Liska-PhD-Prop}  Query relaxation - that is, allowing a query to have terms added, dropped, or re-arranged - has been experimented with and is being optimized. \cite{mias2016, Liska-PhD-Prop}  Another under-experimentation development is operator grouping: the addition and subtraction operators can be classified as one ``type'' of operator, and allowed to be substituted for one another freely; for example, this would let $x + y$ ``exactly'' match $x - y$, $-x + y$, and $-x - y$. \cite{mias2016}  Wildcard inclusion is also planned to be investigated \cite{Liska-PhD-Prop}, and machine learning algorithms will be looked at for use in ranking and disambiguating equations \cite{mias2016}.

To include drastically more math understanding in the engine's procedures, the developers also plan to investigate the inclusion of \emph{Mathematica}-like systems (a.k.a. ``Computer Algebra Systems'') in their mathematics parsing stages. \cite{Liska-PhD-Prop}  Their successful incorporation would transform MIaS into an engine which can translate math - and find matching and similar equations based on that - on par with Wolfram Alpha. This approach has already been partially implemented in an under-development search engine (geared towards physicists), as is discussed in section \ref{sec:5e-descrip}. \cite{Oviedo-Th}

\subsubsection{LaTeXSearch}
Little information is available about LaTeXSearch's algorithms, for its search over some of Springer's journal articles. \cite{LatexSearchAbout}  Independent experiments by this author and  \cite{Liska-PhD-Prop} indicated that it performs presentation-based searching, retrieving both exact matches and similar formulae to a queried equation.

\subsubsection{Uniquation}
Very little of Uniquation's workings is published, and that only on Twitter and the Uniquation web-interface.  Its search over math forums appears to make use of presentation-based search, and finds both similar equations and exact matches. \cite{uniquationWeb}  Basic abstract algebra is supported, such as $\mathbb{Z}[t][\sqrt{t^{2}-1}]$. \cite{uniquation-twitter}  Modular arithmetic is also accessible, including $\sum_{n=1}^m n^m \equiv 1 \pmod p$.  \cite{uniquation-twitter} (These two search queries are presented in section \ref{sec:uniquation-screenshots} as screenshots of Uniquation's search results.  Their LaTeX encodings are also included therein.)

\section{Future Developments}

Developers in the field of math-language search engines have identified a few directions for their development, and a few neat ways in which to expand upon their abilities.

Ideas of this author's, which are specific to physics applications, are also included for discussion and/or development.

\subsection{Field's identified directions}

\subsubsection{Algorithms for Search}
Field experts have indicated that more of the search engines need to focus on extracting mathematical meaning (including context and symbols' meanings) from documents. \cite{AizawaEtAl2014} Algorithms with which to extract theorems from publications, and to restrict searches to formulas which are consistent with a query's mathematical assumptions, are also desired. \cite{MWS-05-2012}  It is being explored to allow the user to define which parts of a formula should be kept fixed, when searching for results, and which are allowed to bring up ``similar'' results (see ``simto'' regions, \cite{ntcir12mathIR}).  There is also work being done to improve the identification of equations which share the same mathematical meaning, but are written differently. \cite{Liska-PhD-Prop, Oviedo-Th} Precision of results and recall speeds are other areas of improvement; the recall speed is only a few seconds at most, but that is still slower than standard text-based search engines.  \cite{Guidi-Cohen-2015} Unbiased evaluation metrics are also recommended, to allow rigorous comparison of the math-aware search engines. \cite{Guidi-Cohen-2015} Finally, scalability is a target area of advancement, with the eventual goal of covering the entire web. \cite{Tangent-pub-2015}

\subsubsection{Database expansion}

Future databases are planned to include the arXiv \cite{liskaEtAl2015ComboTF, MathWebSearch-info}, the digital math library DML-CZ \cite{WebMIaS-descr-pub-2014}, and other added journal articles \cite{WebMIaS-descr-pub-2014, RakosnikEtAl2014}.  The DLMF is also exploring ways to improve or enlarge their special function content (for which they are soliciting community feedback - see ``news'' section of \cite{DLMF-website}). Formalized math libraries (e.g. Mizar) and theorem proving libraries (e.g. Thousands of Problems for Theorem Proving) are also having math-aware search added to them. \cite{MWS-05-2012}

Less formal media which are in-development for math-aware search include lecture videos \cite{Davila2016SIGIR}, Wikipedia \cite{liskaEtAl2015ComboTF, MathWebSearch-info}, and additional math forum websites (such as the question-and-answer forum Math Overflow) \cite{nguyenEtAl2012}.   In addition, an idea has been put forward to use math-aware search on input code for systems such as \emph{Mathematica}, MatLab, and R. \cite{MWS-05-2012}

Eventually, coverage of all mathematical science material is desired \cite{Kohlhase-Scaling-MWS}, with database sizes expected to rival present-day keyword-based search websites \cite{Tangent-pub-2015}.

\subsubsection{User experience}
The user-experience is another focus in development.  This includes re-visiting the subjective rules for which results are presented at the top of the results list, as well as the appearance of the results. \cite{AizawaEtAl2014} Other considerations include ease of use and learning curve. \cite{AizawaEtAl2014}

\subsection{New tools proposed by developers}

Developers have proposed a few tools which use the math-aware search engines, or similar algorithms.

\subsubsection{Alerts for new publications containing a specific equation}
Routing of results - that is, updating users when new documents are available which includes their equation query - is one proposed use of math-aware search engines, by \cite{Youssef-Roles}.  When journal articles comprise the database, this would mean that researchers can set up ``alerts'' to be sent to them when a new article with a certain equation is published (and is indexed into the search system's database).

\subsubsection{New kind of search: Use Diagrams}
The group \cite{Tangent-pub-2015} want to enable search queries to contain diagrams, as well as text and equations.

\subsection{For Physics: Ideas for Useful Directions}

\subsubsection{Rank according to Solvability}
When using these search engines to solve equations, it would be a big help if results were ranked according to whether the equation therein had been solved (exactly or approximately), and how quick it is to solve (computationally or by a skilled mathematician).

Doing so would require first obtaining this information, likely by adding a step to the pre-processing of documents in these databases.  One possible, but probably time-intensive, way to do so would simply be to attempt to solve every one of the database documents' equations in Mathematica, or a similar program.  However, the choice of input parameters would need to be decided upon.  Even an approximate measure of how solvable an equation is would - in general - be better than having a list of unranked equation search results.  Another possible method is to try to extract the information about an equation's solvability from the text or equations of any document which contains that exact written form of the equation; this includes searching mathematical encyclopedias, educational resources, and any other available documents.  This, too, would require a lot of development to be fully implemented.

\subsubsection{Downloading results for post-processing}
None of the search engines allow users to download search results, so that post-processing of equations may occur.

Reasons a researcher might want to post-process equations include:
\begin{itemize}
\item	Input retrieved equations to Mathematica, to see if their written format is easier to solve than the query's written form.
\item	Pattern-search, algorithmically, within the results.  For example, this could help researchers to identify phenomenological patterns, similarities, and analogies.  To implement this, machine learning and data science algorithms may be useful: experimental results and theoretical predictions can both count as ``data'' for these algorithms.
\item	Computational speedup of code: see next proposed application.
\end{itemize}

\subsubsection{Computational speedup: replace slow equations as they occur}
\label{sssec:comp-speedup-future-dev}
Some computations can be done analytically, using Mathematica or similar software.  (Which ones can be done this way is limited by the computation time and ability of the software platform.)  Computations which are done this way often produce secondary equations during run-time: that is, they combine or solve a set of equations, then try to combine it with a second set of equations.  One of the secondary equations can slow down the program's run-time enough that it becomes impractical.  If that equation can be replaced, during run-time, with a faster-solving rewriting of it, then the problem of its slow-down can be fixed.  That replacement could be done by making a run-time call to a ``search engine'' for math equations, which obtains a faster-solving mathematical representation for the problematic equation.

To implement this application, one could place a maximum time on computations.  If the time is exceeded, the math-language search engine is called from the computer code itself.  With no human interaction, the search engine would return a list of equations which match the mathematical meaning of the problematic equation.  A (library) function could sort through the equation results, finding one which lets the code run within a reasonable amount of time.   Preferably, these equations would be ranked by their solvability (as proposed above); however, if they were not sorted that way, the code could iteratively try various of the equivalent equations until either one works or a maximum time limit is reached.

Note that this application requires one or both of the above proposed applications to be done.

\subsubsection{Algorithmically search for patterns, to help direct new research}

Already done in Mathematics, this application involves algorithmically predicting ``probable links'' between known concepts using their mathematical relations, axioms, etc.  These links can then be investigated, and used to help direct new cutting-edge research. \cite{NevzorovaEtAl2014} 

In physics, we could algorithmically find phenomenological patterns, similarities, and analogies.  Considerations would need to include the experimental uncertainty in each experiment, context of equations (for example, does $x$ represent a bead's position on a wire, or a straight distance from an object?), and assumptions behind each experiment and theory which we apply our pattern-discovery algorithms to.  Although these challenges may be considerable to overcome, performing pattern-discovery research without having yet solved them can still provide useful pointers as to where beneficial areas of research might be.

Math-aware search engines' algorithms for matching equations based on their mathematical meaning - both now and as they continue to develop - can provide a good basis for porting this application to physics.

\chapter{Guide to Use}
\label{chap:guide-to-use}


Just as searching Google is a bit of an art, searching by equation requires some training.  The tips given herein may be useful to researchers, when learning the use of these math-aware search engines.  Of course, simply picking an engine at random and inputting an equation will often provide good results, without any of the knowledge elucidated here.

Troubleshooting, wildcards use, tips for successful queries, and algorithm choice are all discussed here. Also included is a short list of tools which help generate LaTeX or MathML equation encodings, as well as tricks for re-using (and navigating) results lists.

\section{Choosing which Engine to Use}
\label{sec:choose-engine}

Equation-based search engines vary a lot in their search algorithms \cite{Oviedo-Th}, database coverage, and results' presentation style.  Yet, no one engine is ``clearly'' better than the rest, according to \cite{Guidi-Cohen-2015}.

Features of each web-available search engine are summarized in tables \ref{table:Web-available-main} and \ref{table:web-available-query}.  Discussing what they mean, and how they might impact a search's goal, are the subjects of this section.

Important considerations for a purpose-driven (or exploratory) search may include: \linebreak whether the search algorithm matches the meaning or the presentation of the math; what fields and types of documents it searches over; and how well you can query given the interface's allowed options.

\subsection{Presentation vs. Meaning-Based Search}
\label{sec:pres-vs-meaning-usage}

Both meaning-based and presentation-based searches have their uses in search, according to developers and this author's analysis.

All meaning-based search algorithms are incomplete, so far: none of them can extract full mathematical meaning from the text around the equation in order to fully use the equation's proper meaning. \cite{MWS-05-2012, Pres-vs-Content-2014}  This is a targeted area of development for the field. \cite{AizawaEtAl2014}  However, some very good progress into meaning-based searching has been made. \cite{MWS-05-2012, AizawaEtAl2014}  Presentation-based searching also has developments underway, surrounding how best to find ``similar'' equations (the meaning of which is naturally subjective) to an input query. \cite{ZY_math-sim-search}

\subsubsection{Definitions}
Presentation-based searching attempts to match the visual structure of equations \cite{DLMF-AMS-notice, Liska-PhD-Prop, Tangent-technical-v0.3}, whereas meaning-based algorithms work to account for equivalencies of mathematical meaning between differently-structured equations \cite{Pres-vs-Content-2014, Liska-PhD-Prop, MWS-05-2012}.  For example, $(x^2 + y^2)/x^2$ has the same meaning as does $1 + (y^2/x^2)$, but is not visually equivalent.  The two equations presented here would match in a meaning-based search, but not in a presentation-based search.

\subsubsection{Uses of Each}
Presentation-based search enables equations which are found to be written in similar notation and/or format to the one which was input.  Two studies \cite{Tangent-pub-2015, Landy-Goldstone-2007, ReichenbachEtAl2014} have suggested that these properties - notation and format - play a role in how individuals reason about math.

If the true physical phenomena are sought after, however, meaning-based math matching is required, for it recognizes that math is just a representation of physical or abstract systems.  To describe a system, one can use many different math notations: the system does not care which one you use, it will behave the same either way.

As such, presentation-based search is good when users seek familiar notation.  For example, it would be useful to look up information related to a formula which is fairly standardized, to understand an unfamiliar notation of a formula from a document \cite{Tangent-pub-2015,Wangari-Th}, and to find out about specific questions or formulae in a given branch of math \cite{Wangari-et-al_2014_paper}.  When a person does not know the name of a type of math, or of a specific formula, then presentation-based search also provides great usefulness.

Math-meaning searches complement presentation-based search.  Formulas found from meaning-based searches can be found in presentation-based search to enable better understanding of the math therein.  On the other side of the coin, presentation-based searching can provide better context and refine a search, if that is desired, before using meaning-based search.  In some cases, of course, only one or the other will be needed.

Meaning-based searches, alone, can hit directly upon the deep connections which math notation hides.  In this sense, they are far more useful for branching between fields of physics and/or math, and for (eventually) algorithmically searching for patterns between the phenomena which math represents.  They help researchers to gain better understanding of what the system is actually doing, by changing the human construct imposed upon it by the math: this can help the researcher to separate the system from the mathematical formulations.

\subsubsection{Complexity of equation: A consideration}
A small experiment \cite{Pres-vs-Content-2014} found that presentation-based search engines provided better results than did meaning-based engines for the following types of equations: elementary functions (such as logarithm), and equations which have only one way of being written (excluding, for example, certain trigonometric expressions).  Special functions which have a standardized representation, such as the Legendre-Q function, were suggested to be equally well-matched by presentation-based and meaning-based search engines.  Equations which have multiple written forms were found to be better-matched by meaning-based search engines; examples include trigonometric expressions and certain special functions (e.g. Hermite-H, Poly-Gamma).

\subsection{Documents Searched: Field and Document Type}

Different databases are searched by each search engine.  Documents include journal articles, mathematics' special functions databases, and Wikipedia.  Fields of research covered include physics, astronomy, applied mathematics, statistics, pure math, and others; each of the ones named here is especially applicable to physics research.

\subsection{Query Features}

\subsubsection{Wildcards}
\label{sssec:wildcards}
Some engines allow wildcard searches.  Which ones do, and their wildcard characters, is given in table \ref{table:web-available-query}.

Named wildcards are allowed by some searches.  They can be made, for example, by preceding a character with a ``?'', such as ``?a''.  If the same named wildcard occurs more than once in a query, then it enforces that results should have the same variable in that wildcard's place; for example, ``?a + ?b\textasciicircum\{?a\}'' could return $x + y^x$ but not $x + y^z$.  Expressions, single variables, or constants can be allowed to take the place of a named wildcard. \cite{ntcir-10-overview, ntcir-10-mws, zbMATH-engine}

In the literature, ``query variables'' is another term which is used to describe wildcards. \cite{ntcir-10-overview, Liska-PhD-Prop}

\subsubsection{Mutli-equation queries: Initial conditions inclusion}
Some search engines allow the user to search for multiple equations at once. \cite{SojkaEtAl2011, Tangent-article-2015, Tangent-pub-2015, DLMF-website}  This allows initial conditions to be added to a search, among other things.  The searched equations do not have to occur in the same portion of the paper.  As with Google, these equation-based search engines may relax the query if the equations can not be found to occur in the same document: as does MIaS \cite{SojkaEtAl2011}, they might require only one or more of the multiple query equations to be found in the paper.

\subsubsection{Author, DOI, \& other searches alongside equations}
In many of these equation-based search interfaces, queries can include both equations and keywords from a specified portion of the publication.  Author, year, journal, DOI, and title are common options.

\section{Successful Queries}
\label{sec:successful-queries}

Successful and efficient querying is the goal of this section.

\subsection{Making the query}

\subsubsection{Copy and paste an equation from a publication}
If the equation query is a complex equation from an existing publication, then it is sometimes possible to access its LaTeX code.  Springer articles can be found via LaTeXSearch, with their equations listed for easy review; however, this database contains a limited number of Springer's articles, so finding the equation is hit-and-miss. \cite{Springer-LatexSearch-post, LatexSearchAbout, Springer-main-page} Another route is to download the LaTeX source code from arXiv.org, if the publication is available there.

\subsubsection{Download code for a special function}
If a special function is your query, it might be found in the DLMF \cite{DLMF-website} or Wolfram Functions website \cite{WolframFunctionsWebsite}.  Both are searchable by keyword (and the DLMF is also searchable by portions of equations).

To download an equation's code in DLMF, click the equation's chapter/section link in the results list  This will lead to a page with more information, and many equations.  Next to each, there is an encircled ``i'': click it (do not just mouse-over it).  Select ``TeX'' to download a TeX file with the equation in LaTeX code.  For MathML, click ``pMML'' (``presentation MathML'') to open a new webpage with the code. \cite{DLMF-website}

Wolfram Functions allows code to be downloaded in MathML, by going to a specific formula's webpage therein.  For instance, one can search for ``gamma function'', click the results titled ``gamma function'', and in the list of its (193) formulas, click on ``primary definition''.  On the page which comes up, click the equation.  The resulting webpage has MathML code for the formula, as well as \emph{Mathematica} code. \cite{WolframFunctionsWebsite}

\subsubsection{Generate equation code}
\label{sec:template-editors}
Input of an equation query is usually done in LaTeX or MathML.  Generating code in either language is simple with the aid of online template-based editors.  Drawing an equation or uploading its image are both possibilities \cite{min_Sasarak-et-al_2012}, although image upload is currently being fixed. \cite{min-developer-email}  Translators also exist between the two languages.

Examples of tools are given here; they create or translate LaTeX or MathML code for equations.

\paragraph{Draw equation}
The online editor $m_{in}$ allows users to draw equations, then send them as queries to various search engines or download their code as TeX or MathML. \cite{min_Sasarak-et-al_2012} To copy the code for a drawn expression, right-click on the expression's image in the small window near the top right corner of the screen; a drop-down menu with ``save math as'' will appear. \cite{min-video}

\paragraph{Upload equation's image}
LaTeX code can be generated from an image or PDF file of an equation.  This is not only useful for grabbing equations from academic papers, but also for extracting equations from online services such as the Wolfram (special) Functions website.

A desktop software exists to do this, which translates TIF, GIF, PNG, BMP, or PDF files to LaTeX or MathML. \cite{InftyReaderTextFile}  It can translate entire documents, or single-equation images. \cite{InftyReaderDownload}  The free demo version is fully featured, and allows up to 5 pages of documents per day. \cite{InftyProjectAbout}  The software is called InftyReader. \cite{InftyReaderDownload}

The web-hosted tool $m_{in}$, which has been discussed elsewhere in this document, also allows images to be uploaded for translation to LaTeX; however, that feature is under repair at the moment, and is expected to be fixed in late 2016 \cite{min-developer-email}.  Equations which are uploaded in this way enter into $m_{in}$'s drawing board, where they can be edited.  The final equation can then be sent to various math-aware search engines as a query, or its LaTeX code can be downlaoded. \cite{min_Sasarak-et-al_2012, min-video}

\paragraph{Template editors}
\textbf{LaTeX:} At least a few template-based editors are available which generate LaTeX code from a template editor.  One online version is Springer's LaTeX Sandbox. \cite{Latex-sandbox}
\textbf{MathML:} A good list of equation editors and viewers (``renderers''), for MathML, is provided at \cite{MathML-tools}.  Equation editors include downloadable interfaces (such as fMath and MathMagic) and plugins (such as FireMath, a Firefox plugin). \cite{MathML-tools}  A good renderer for MathML equations, which generates a math equation image from the code, is Wolfram Research's MathML renderer. \cite{MathMLrenderWolfram}
\textbf{Both LaTeX and MathML:} For licensed or subscription users, \emph{Mathematica} and Wolfram Alpha Pro each allow export of formulas to MathML or LaTeX code.  In \emph{Mathematica}, exporting is done by right-clicking on an output formula and selecting ``Copy As''.  In Wolfram Alpha, a mouse-over on the formula's image brings up a menu; clicking ``data'' brings up a menu from which Pro users can download the formula's code.

\paragraph{Translators}
Translating from MathML to LaTeX can be useful, for example when copying special functions from the Wolfram Functions website.  However, given the limited options available for this, it might be easier to copy the relevant equations' pictures and upload them to a software which translates them into LaTeX.  Although many LaTeX to MathML translators exist online (e.g. MathMLCentral \cite{mathmlcentral}, MathToWeb \cite{mathtoweb}, LaTeX2MathML \cite{latex2mathml}, TeXZilla \cite{texzilla}), no operational web-hosted services for MathML to LaTeX could be found (one example of an inoperative one is “MathML to TeX online Translator “ \cite{orccaMathMLtoLatex}).  There are a couple of desktop solutions, although there seems to be no working desktop software.  One solution is to use the github-hosted program web-XSLT \cite{webxslt}: there are no usage instructions included, but the developer specified on a forum that it should be used with Saxon (or a different xslt2 engine) and run via the command ``java -jar saxon9.jar myfile.xml pmml2tex.xsl'' \cite{webxsltStEx}.  Alternatively, the XSLT MathML Library can be used with Saxon, via ``saxon -o output.tex input.mml xsltml\_2.1.2/mmltex.xsl''. \cite{xsltMathMLlibraryForTex}

\subsection{Wildcards}

Some search engines allow wildcards.  This is discussed in section \ref{sssec:wildcards}.  Each search engine's treatment of wildcards is summarized in table \ref{table:web-available-query}.

\subsection{Troubleshooting}

If an equation query returns no results, or unsatisfactory ones, then the following might help target results better.
\begin{itemize}

\item	Check whether this math-aware search engine requires equations to be bracketed by TeX \$s.
\begin{itemize}
\item	For instance, it might require the equation to be entered as ``\$\$equation\$\$'', ``\$equation\$'',  or ``equation'' (without the quotation marks).
\end{itemize}

\item	Change the notation: \cite{Guidi-Cohen-2015} point out that non-standard math notation can make a query hard to locate.
\begin{itemize}
\item	For example, some authors use ``.'' (which appears as ``$.$'') or ``\textbackslash bullet'' (``$\bullet$'') to indicate the dot product, instead of ``\textbackslash cdot'' ($\cdot$).  Different LaTeX encodings are accounted for to a certain extent by the math-aware search engines, but this type of non-standard notation can provide a problem; this is especially pertinent for ``.'', which is interpreted as a decimal point.  Therefore, it can be useful to try using both standard and non-standard LaTeX notation in the query. \cite{Liska-Th}
\end{itemize}

\item	Simplify or otherwise re-write the formula.
\begin{itemize}
\item	For instance, the formula $\sinh(\log(\cos(x)))$ simplifies to \linebreak $(-1/2)\sin(x)\tan(x)$ (and a few other forms) \cite{Wolfram-Alpha-web}, which would likely provide different search results.  As a simpler example, searching for $a\sin(a) + a\cos(a)$ can provide different results to a query of $a(\sin(a)+\cos(a))$.  These rewritings are only necessary because the math-aware search engines don't yet have the ability to compare equations based on their meaning in math; the next generation of engines should ``know'' all of the same math rules as Computer Algebra Systems (such as \emph{Mathematica}), and are under development already \cite{Oviedo-Th, Liska-PhD-Prop}.
\end{itemize}

\item	Generalize the formula.
\begin{itemize}
\item	For example, one can transform elementary functions into simple variables (e.g. $\sin(2x)$ into $f(x)$) or exchange an expression in a query for a variable (e.g. rewrite $\sin(\log(\sqrt{x+2})$ to $\sin(\log(a))$).
\end{itemize}

\item	Add a keyword to the search.
\begin{itemize}
\item	For example, to find information about how the ``prime counting function'' implies that $\pi(2) = 1$, it can be useful to search for both $\pi(2) = 1$ and ``prime counting function'' in the same query. \cite{Wangari-et-al_2014_paper}
\end{itemize}

\end{itemize}

\section{Results}

Search-easing features of some search engines, effect of different algorithms for ranking results, and expectations for display of results are included in this section.

\subsection{Easing the search, in results stage}

\subsubsection{Result as a new query}
The LaTeX text of results can be accessed in LaTeXSearch, making re-use of a result easy. \cite{LatexSearchAbout}  In addition, $m_{in}$ allows images of equations to be used as search queries \cite{Min-instructions}, making it much easier to re-use results from any search engine's query;  however, this feature is currently in-operational, and is expected to be fixed shortly \cite{min-developer-email}.  Tangent provides a direct link from each result to open its equation in $m_{in}$ (section \ref{sec:min-description}), where users can edit it via drawing or text, then generate LaTeX or MathML code. \cite{min-video}

\subsubsection{Filter results}
Results can be filtered or sorted by publication year, author, journal, and other subjects, in some search interfaces.

\subsection{Ranking Algorithms' Effects}
Results rankings differ between engines.  Since different rankings are appropriate to certain fields of study \cite{SojkaEtAl2011, Rec-and-retrieval-rev} and goals of search, it may be beneficial to both try a few different engines and skim a few pages of results.

\subsection{Presentation Expectations}
Most of the search engines display results very well, including a display of the matching equation, text for context, a link to the article or webpage, DOI, and articles' titles and authors.  Engines which do not display results well have that noted in their summary of section \ref{sec:web-avail-engines}.

\chapter{Available Math-Aware Search Engines}
\label{chap:avail-SEs}

The math-aware search engines which are available include many which are intended for use online, and a few which are downloadable for local creation of a math-searchable database.

Included here are summaries of each engines' abilities, including only those which work on the web or (for downloadable ones) have a working implementation in an existing database.  Their features are summarized, for comparison and quick-reference in tables \ref{table:Web-available-main} to \ref{table:downloadable-querying}.%

Because algorithms and databases differ, it can be useful to search using many math-aware search engines.  To make the search more efficient, a multi-engine interface exists ($m_{in}$). \cite{min-interface}  In addition, it is simple to copy and paste equations from one search engine to another, since they all accept LaTeX input.


\section{Engines Included: Working, Web-Available}
\label{sec:engines-avail}

Many prototypes are in development, and many old architectures are no longer available (or, operational) on the web. \cite{Guidi-Cohen-2015}  Because this document is tailored towards use of the search engines, only working and web-available search engines are included.  Some of these have back-end software which is available for download, for the creation of equation-searchable databases from local or web-crawled documents; these features are also summarized.

One exception to the web-available rule is that a physics-catered search engine for equations is included in this summary.  Its developers (from CERN) have worked to include physics-specific considerations in the search algorithms.  Physics, after all, is not purely math.  The features of this up-and-coming equation-search engine are overviewed.

Architectures which were discovered, but not included due to absent web-availability, are listed here in case the reader comes accross them elsewhere.  They may be in early stages of development, or old enough to be discontinued.  They include: EgoMath2, Ego-Math, Physikerwelt, WikiMirs, Formula Search Engine, MathDex (formerly MathFind), LeActiveMath, ActiveMath, Math Go!, and Math Aware Search Engine (MASE). \cite{SojkaEtAl2011, Oviedo-Th, Liska-Th, Liska-PhD-Prop, WikiMirs-Main-pub, ActiveMath-workings, Wangari-Th, Pres-vs-Content-2014}

\section{Keeping Up-To-Date}

Keywords to help researchers keep abreast of newly-developed math-aware search engines include:
\begin{itemize}
\item Math Information Retrieval (MIR)
\item Mathematics Retrieval
\item Mathematical Knowledge Management (MKM)
\item Digital Mathematics Libraries (DML)
\item Math Search Interfaces
\item Math-Aware Search Engines
\item Formula Search
\end{itemize}

Terminology in this field can vary, as conventions do not seem to be yet set.  For example, the following types of search can vary in their detailed definitions (while matching broad ideas) between documents: ``similarity'', ``presentation''-based, ``syntactic'', ``semantic'', ``content-based'', and ``meaning-based''.

\section{Multi-Engine Interface}

\subsection{$m_{in}$}
\label{sec:min-description}

Four equation-based search services, plus Wikipedia and Google, are searchable from the single input interface $m_{in}$. \cite{Min-instructions}  The math-aware search engines available are:
\begin{itemize}
\item Tangent (searches an old version of Wikipedia)
\item NIST DLMF (provides special functions)
\item LaTeXSearch (searches science and math journals)
\item Wolfram Alpha (generates information, sometimes including special functions).
\end{itemize}
Each selected search is opened in a new tab, so that the original input is preserved for use in another search engine. \cite{min-interface}

Drawing or typing equations is allowed.  Image recognition of png-format equations is also a feature, but is under repair at the moment. \cite{Min-instructions, min-developer-email}


\section{Web-Available Engines}
\label{sec:web-avail-engines}

The engines summarized here have a working online interface, at the time of this writing.  Their features are summarized in tables \ref{table:Web-available-main} and \ref{table:web-available-query}, which are contained on a single page for quick reference.  Example searches can be found in section \ref{sec:math-meaning-limitations}, which outlines the extent of each engines's mathematical search ability, and chapter \ref{chap:screenshots}, which contains screenshot of searches with each engine.

\subsection{LaTeXSearch}

Covering physics and math journal articles, ranging from quantum theory to energy, \linebreak Springer's LaTeXSearch enables presentation-based equation retrieval.  A subset of their articles are included, but Springer does not specify which subset \footnote{Springer specifies that 120,000 peer-reviewed articles are covered by LaTeXSearch.  However, Springer's main webpage states that the size of their entire collection is over 5 million articles. \cite{Springer-LatexSearch-post, LatexSearchAbout, Springer-main-page}}. \cite{LatexSearchAbout, Springer-main-page}

This platform is proprietary, and provides very little information about its internal search algorithms.  Both exact matches and similar equations are found. \cite{LatexSearchAbout}  A ``Speed-search'' feature searches for only exact matches to the input LaTeX code (which is different from searching for exact matches to the equation, as a given equation can have multiple LaTeX encodings).  It runs much faster than does the full LaTeXSearch. \cite{LaTeXSearch-search-page} Based on this author's experimentation, this engines' ``similar equations'' include those which have different variable names, different constant values, or variables in the place of constants (and visa-versa).

Presentation-based search appears to be used, based upon independent experiments by: this author, \cite{SojkaEtAl2011} and \cite{Pres-vs-Content-2014}.

The language of query is LaTeX.  Equation terms which are entered into the query box are all considered to be part of a single equation: multiple equations can not be searched for \footnote{This was found from experimentation.  Specifically, a paper was found which contained at least two equations, and the two equations were searched for independently (to make sure they were found) and then together, in the format (without the \textlangle \  \textrangle \  brackets): \textlangle \ ``equation1'' AND ``equation2'' \  \textrangle \  When they were searched for together, no results were found.  Searching for pieces of a single equation, in the format \textlangle \ ``piece1'' AND ``piece2'' \ \textrangle, does work to bring up individual equations which contain both pieces.}.  Advanced search options allow subject, publication, and year to be added to the equation search. \cite{LaTeXSearch-search-page}  Alternatively, a single article's equations can all be recalled by entering the DOI or title keywords of that publication.  \cite{Springer-LatexSearch-post}  

LaTeX code for the discovered equations is easy to download, and is located below the equation's image in the results. \cite{LaTeXSearch-search-page}

The developers want to expand the database, including possibly indexing the arXiv.  They ask people who are interested in helping to contact them. \cite{LatexSearchAbout}

\subsection{zbMATH}

Natural sciences (including physics), computer science, and other applied math publications comprise a significant portion of the zbMATH database.  Pure math publications make up the rest.  They include articles from around 3000 journals, and 170,000 books, comprising over 3 million publications in total.  Many old texts have been digitized, as well, for inclusion in the database. \cite{zbMath-database-about, RakosnikEtAl2014}

Using MathWebSearch as its engine (detailed in section \ref{sec:MWS-avail-engines}), zbMATH provides \linebreak meaning-based math search.  Exact matches to an equation are found, unless named wildcards are used, in which case similar equations are found.  Named wildcards are made, in this case, by preceding a character with a ``?'', such as ``?a''. \cite{ntcir-11-mws}

A query can include a single ``LaTeX-style'' equation, along with several optional keywords. \cite{zbMATH-engine}

Results are displayed as a list of titles with author and year.  Clicking on one of the results expands it, showing a highlighted form of the located equation, along with the article's abstract, DOI, and other information. \cite{zbMATH-engine}

A subscription service of zbMATH is also available, and is not to be confused with its free equation-search platform.  The subscription service adds other query options, such as author, publication year, and mathematical classification, to an equation-based search.  It can be previewed, but only three results are available in the free version.  Access to it is found by clicking the ``zbMATH'' logo at the top of the formula-search (free) platform, \cite{zbMATH-engine}.

\subsection{EuDML}

Mathematical Physics journal articles, and other math journal articles, are included in EuDML.  Over 250,000 documents are searchable, including a few books. \cite{EuDML-web-base} The developers are negotiating to include more content, but they note that it is difficult to convice ``big commercial publishers'' (such as Elsevier and Springer) to agree to their article control policies.  Other publishers with whom they have been under negotiations with include Eulicd, Math-Ru.Net, and EMS Publishing House. \cite{RakosnikEtAl2014}

EuDML allows a search query to include a single equation and any number of keywords.  Note that, while its engine (MIaS) does allow mutliple equations to be included in one query \cite{SojkaEtAl2011}, experimentation reveals that the EuDML interface does not seem to do so. \cite{EuDML-web-base}.  Other search categories can be specified in EuDML, as well, including author, year, and language of writing.  \cite{EuDML-web-base}

Results' presentation could be improved, but is sufficient: the equation from the result is not highlighted, and is often not displayed in the text of the result.  Finding the equation, and discovering whether it was an exact vs. similar match to the query, requires accessing and downloading the article (or book) which was found to contain a match.  Included in each result is the title, author, year, and usually some text.  \cite{EuDML-web-base}

Presentation-based search is performed; both similar and exact formulae are found.  The back-end engine used, MIaS \cite{Liska-PhD-Prop}, is further described in sections \ref{sec:MIaS-meaning-abilities} and \ref{sec:MIaS-descrip-Avail-engines}.

\subsection{NIST Digital Library of Mathematical Functions (DLMF)}

Special functions from the textbook ``Handbook of Mathematical Functions'' have been digitized and expanded upon to provide an equation-searchable library.  Highly curated, this library consists of formulae, tables, 3D visualization, and graphs. \cite{DLMF-website}

Formulae can be downloaded for their LaTeX or (presentation) MathML versions. \cite{DLMF-website}

Search types possible include text, phrases (in double-quotes), LaTeX-like math expressions, boolean operators, and proximity operators (which denote how close a term should be to another term).  Categories of math, such as ``trigonometric'' can be used in the search.  \cite{DLMF-help}  Pieces of many equations, such as $\langle \text{\ (1-z\textasciicircum{}2)\textasciicircum{}(1/2) and e\textasciicircum{}\{nu(1-x)\}} \ \rangle$, can be searched for by connecting expressions with $\langle \text{\ and} \ \rangle$; the query's exact equations, or their similar counterparts (see next paragraph) must all appear in the same section of the online textbook in order for it to be retrieved as a result. \cite{DLMF-website}

Presentation-based search is used by DLMF.  First, the search engine attempts to find an exact match to the input equation.  After those have been found, the symbols of the equation are allowed to be re-ordered or separated by other math terms.  \cite{DLMF-AMS-notice, Pres-vs-Content-2014}  As a consequence, this search engine can not identify special functions which are written in a way which significantly differs from that of its internal database.  However, it can still be used to look up some representations of special functions.
Software by which to solve special functions is also listed in the DLMF, including a table specifying which software solves specific special functions. \cite{DLMF-Software-Index}

\subsection{Tangent}
\label{sec:Tangent-web-avail}

A section of Wikipedia is the database for this search engine.  The content is frozen to a certain date: it does not update as Wikipedia evolves. \cite{dprl-descriptions}

Only ten results are allowed to be viewed, in this demo. \cite{TangentSE} It is not clear why; however, this may often be enough to find the desired results on Wikipedia.

Presentation-based search is used, which the developers say helps non-specialists find understandable information. \cite{Tangent-pub-2015}  Visual aspects, including left-to-right structure, are considered when ranking results. \cite{Tangent-technical-v0.3, tangent2016sigir, TangentSE}  Exact matches are listed first; similar matches might have variables' names changed, constants in place of variables (or visa-versa), constants' values changed, or changes in the structure of the equation. \cite{Tangent-pub-2015, TangentSE}

Users can search for multiple equations at once. \cite{TangentSE}  While keywords can be added to the query, the results may be unexpected since they are not supported in this web-available version. \cite{Tangent-technical-v0.3, TangentSE}  Input is allowed via LaTeX or a drawing interface (named $m_{in}$). \cite{Tangent-article-2015, Tangent-pub-2015} Wildcard use is supported, with the ``?'' symbol representing any single symbol. \cite{Tangent-pub-2015}

A downloadable, more advanced version of Tangent is available for enabling math-search over custom databases.  It is accompanied by clear instructions as to how to run a webserver which searches a website for equations.  (See section \ref{sec:Tangent-avail-engines-download} for more details.)

\subsection{Uniquation}

This engine searches over math ``help'', or ``question and answer'' websites, such as planetmath.org, mathoverflow.net, and math.stackexchange.com.

Not much information is available about it.  No scientific papers can be found which focus on it, although it is mentioned in a few papers (such as \cite{Oviedo-Th}).

It appears to use presentation-based search: it finds structurally similar and equivalent expressions.  Variable names are ignored, when searching.

TeX is used as the language of query.  Keyword-searching does not appear to be supported.

Results are presented grouped together by formula; the actual webpages are accessed using a link above each formula. \cite{uniquationWeb}

\subsection{Wolfram Alpha}

Wolfram Alpha both retrieves mathematical (and other) information from an encyclopediac database, and calculates results to queries.  It has roughly 1000 curated data sets, and draws from real-time data sources.  \cite{Weisstein-Alpha} Many special functions can be found, although it does not recognize all forms of the special functions (see \ref{sec:Wolfram-Alpha-Math-Meaning-Extent}); definitions, information, and plots about functions are also presented.  \cite{Wolfram-Alpha-web}  It can process math based on its mathematical meaning, due to its use of the math processing power of \emph{Mathematica}. \cite{Weisstein-Alpha}

Wolfram Alpha does not bring up instances of formulas in journal articles or webpages, but rather brings up information from its own internal curated database; in this sense, it is more like an encyclopedia which is accessed solely via search queries. \cite{Wolfram-Alpha-web}

Input can be given in natural language \cite{Weisstein-Alpha} (English \cite{WAfaqs}), TeX \cite{WAunderstandsTex}, and (the majority of) small pieces of \emph{Mathematica} code \cite{WAfaqs}.

\section{Ready for ArXiV deployment: Needs Server Space}

The developer of the equation-based search engine MathWebSearch, Michael Kohlhase, states on his webpage that MathWebSearch is ready for deployment on arXiv.org.  To do so, he needs 128GB of additional server space, and that their restriction comes especially from the RAM; donations of server space are invited. \cite{MathWebSearch-info}

MathWebSearch currently powers zbMATH, and is available for download to build local and web-crawled searchable databases. \cite{MathWebSearch-info}  It also has a demo interface at \cite{Sentido-search}, but that demo is only partially operational at this time.

\section{In development for physics}

\subsection{$5e^{x+y}$}
\label{sec:5e-descrip}

CERN has recently been developing a physics-catered search engine for equations.  It is meant to be used on journal articles and other scientific documents, and is currently catered to the CERN Document Server (CDS). \cite{Oviedo-Th}

Meaning-based search is used: equations are parsed for comparisons using \emph{Mathematica}.  The ``Simplify'' and ``FullSimplify'' commands, in Mathmatica, are used to normalize every equation in the database, and every query, so that they can be compared based on their mathematical meaning rather than their syntax (or ``presentation''). \cite{Oviedo-Th, Oviedo-5e-2pg}.  These commands in \emph{Mathematica} can re-arrange the equations to a standardized form, using special function definitions, elementary function definitions, and of course simple rules including associativity and distributivity (which the other search engines only approximate).  \footnote{Another, currently-available math-aware search engine is investigating including \emph{Mathematica}-like systems for performing the mathematical comparisons: MIaS, which has its present state discussed in sections \ref{sec:MIaS-meaning-abilities} and \ref{sec:MIaS-descrip-Avail-engines}. \cite{Liska-PhD-Prop}}

Physics considerations are also in the forefront.  Query constants' values are included in the search in a way which accounts for different experiments of the same phenomenon giving different results: their value is iteratively rounded, for matching purposes.  Mathematical structure is emphasized in the search.  Specific fields of study have common patterns, such as $ \langle \  S = \int_{t1}^{t2} \mathcal{L} dt \  \rangle$ or $ \langle \text{ originalParticle}$ $\rightarrow$ $\text{subparticle1,}$ $\text{subparticle2,}$ $\text{subparticle3 } \rangle $; many of these were manually added to the search algorithms, although the list is not meant to be exhaustive. \cite{Oviedo-Th}  The developers also tried to account for the fact that variables' names follow certain conventions in some cases, and are irrelevant in others. \cite{Oviedo-5e-2pg}

More development is needed before it is ready for full-scale deployment on CDS and similar systems. \cite{Oviedo-Th}  However, a working demo version was available on Invenio's main webpage \cite{Oviedo-5e-2pg}; it has been taken down now \cite{Oviedo-PC}.  Hopefully work will continue on this physics-catered project.

\section{Build a Custom Equation-Searchable Database}

Universities, institutions, and individuals can enable math-equation searching on custom databases of MathML (using MWS, Tangent, MIaS) and LaTeX (using Tangent).  Web-crawling (using MWS) or local databases (using Tangent, MIaS, MWS) can be used.

The software for this also powers the web-available engines.  Other software might exist for this, which has no working web demo or interface; those are not included here.

Features of the included engines are summarized in tables \ref{table:downloadable-main} and \ref{table:downloadable-querying}.  More details about the mathematical processing abilities of these engines is given in section \ref{sec:tangent-extent-abilities}.

\subsection{Tangent}
\label{sec:Tangent-avail-engines-download}

More advanced than its web demo, which is v0.1, Tangent v0.3.1 is available for download from the research group's webpage: \cite{dprl-descriptions}.

This updated version is the latest release under development by \cite{dprl-descriptions}.  It finds similar equations to a query, based on their symbols and layout. \cite{tangent2016sigir}  Keywords are supported in the same query as equations, and wildcards can represent subexpressions or single symbols.  \cite{ntcir12tangent}  This scalable \cite{tangent2016sigir} engine can process local documents in TeX and MathML \cite{Tangent-thesis, DPRL-github}, and has been used on parts of the arXiv and Wikipedia \cite{ntcir12tangent}.

\subsection{MathWebSearch (MWS)}
\label{sec:MWS-avail-engines}

Providing meaning-based search \cite{MWS-05-2012, SojkaEtAl2011, Pres-vs-Content-2014, Liska-PhD-Prop}, this scalable engine can work independently \cite{MWS-05-2012} or with other ``wrapper'' software which adapt it to different database types, user interfaces, and ranking schemes \cite{MWS-wiki}.  A version for the arXiv has been tested \cite{MWS-05-2012}, and needs only server space to fully implement \cite{MathWebSearch-info}.

MWS uses its own algorithms to provide meaning-based search, and is based on Lucene for text-based searching. \cite{SojkaEtAl2011}  It supports named wildcards, so that a character can be replaced by any expression or single variable in results; they are denoted by preceding a character with a ``?''. \cite{zbMATH-engine, ntcir-10-overview, ntcir-10-mws} A description of named wildcards is given in section \ref{sssec:wildcards}.

A working search interface which uses MWS is zbMATH. \cite{zbMATH-engine}  Another demo, called Sentido-search (\cite{Sentido-search}), showcases an MWS input panel which both has a template and allows for and coding in many languages, including LaTeX; it has a search feature, but it is currently not fully functional.

ArXiv.org deployment of MWS is stated to need only an additional 128GB of RAM; donations of server space are invited. \cite{MathWebSearch-info}  MWS developers tested its performance on about 65\% of the arXiv (as of 2012).  Query times averaged 25 - 75 ms for most queries, and went just past 300 ms for a more difficult class of query.  The system uses distributed computing, so that its estimated memory requirements for the entire arXiv (170GB) can be accommodated.  A user interface for arXiv retrieval was designed, and appears to use LaTeX for its query language. \cite{MWS-05-2012}

The github version of MWS allows indexing of MathML formulae from a website or a local repository of XHTML documents. \cite{MWSgithub}  (To use it on arXiv.org documents, the developers use an XML-translation of the arXiv.org files. \cite{MWS-05-2012, XML-arxiv})  Instructions for installation of MWS are included on the github page (\cite{MWSgithub}).

\subsection{Math Indexer and Searcher (MIaS)}
\label{sec:MIaS-descrip-Avail-engines}

MIaS uses presentation-based searching; upgrades to meaning-based search, with the help of \emph{Mathematica}-like systems, is being explored. \cite{Pres-vs-Content-2014, Liska-PhD-Prop, SojkaEtAl2011, Liska-Th}  First locating queries' exact matches, in the database, it then moves on to allow variable names to be ignored (unless the query consists of solely a single variable), constants' values to not matter, and only a portion of the formulae (i.e. their subformulae) to match.  New developments also allow it to match equations which have similar visual structure. \cite{mias2016}  The complexity of both the query and database equations are considered when ordering the retrieved formulae: in general, more complex formulae are ranked higher in the list of results. \cite{SojkaEtAl2011}

Queries allowed include multiple equations, alongside textual keywords.  When multiple equations are searched for, only one needs to appear in the results: the equations are connected internally by the OR operator. \cite{SojkaEtAl2011} Input is done, at present, in LaTeX or MathML; a visual rendering of the input is provided. \cite{SojkaEtAl2011}  Wildcards are currently being explored for inclusion. \cite{Liska-PhD-Prop}

Already used in EuDML, this back-end engine is designed to support large-scale libraries such as arXiv. \cite{SojkaEtAl2011, Liska-PhD-Prop}  It runs on the platform Apache Lucene \cite{SojkaEtAl2011}, which is already in use for the CERN internal document server (CDS) \cite{Oviedo-Th}. 

A demo of the downloadable interface, WebMIaS, is provided at \cite{MIaS-Demo} and described in \cite{WebMIaS-descr-pub-2014}.  The demo searches around 100,000 arXiv.org articles (the ``NTCIR-12'' dataset). \cite{WebMIaS-descr-pub-2014, WebMIaS-webpage, NTCIR-12-database}  Input is either TeX or MathML, and keyword-search in the ``title'' or ``content'' can accompany an equation in a query.  Presentation- and/or content- MathML language can be used: the former can be better for visually-matching equations, and the latter for matching their meaning. \cite{WebMIaS-descr-pub-2014}  The results display well, with formulae matches highlighted.  Unfortunately, at present, the documents from which results are drawn are inaccessible from the webpage; however, arXiv numbers can often be recognized in their titles, so the results can be found by the determined user. \cite{MIaS-Demo}

The github versions of MIaS \cite{MIaS-github} and WebMIaS \cite{WebMIaS-github, WebMIaS-descr-pub-2014} use Lucene to index and search a local database of files; math in the files must be written in MathML. \footnote{For arXiv-interested users, note that a partial version of the arXiv.org documents is available in MathML: it is called arXMLiv. \cite{XML-arxiv}}  A plug-in to Solr and Lucene is also available, called MIaSMath. \cite{MIaSMath-github}  Instructions for use are provided on the github webpages (\cite{MIaS-github, MIaSMath-github}).

\newpage

\begin{landscape}

\begin{table}[H]
\begin{threeparttable}
\caption{Web-Available Search Engines for Math}
\small
\centering
\label{table:Web-available-main}
\begin{tabular}{SlSlSlSlSlSl}  \toprule %
Search Engine & Discipline\tnote{A} & Database & Search type & Back-end & Web address\\ \midrule 
\rowcolor[gray]{.9} %
LaTeXSearch%
&Math-Sci%
&Journal articles%
&Presentation%
&Custom%
&\footnotesize{http://latexsearch.com/home.do}\\ %
\rowcolor[gray]{.9}%
&Math%
&%
&%
&%
&\\
zbMATH%
&Math-Sci%
&Journal articles,%
&Meaning%
&MWS%
&\footnotesize{http://search.mathweb.org/zbl}\\  %
&Math%
&Books%
& %
& %
& \\%
\rowcolor[gray]{.9} %
EuDML%
&Math-Sci%
&Journal articles,%
&Presentation%
&MIaS%
&\footnotesize{https://eudml.org/} \\ %
\rowcolor[gray]{.9} %
&Math%
&Books%
& %
& %
& \\ %
DLMF%
&Math%
&Handbook of Mathematical Functions%
&Presentation%
&Custom%
&\footnotesize{http://dlmf.nist.gov/}\\ %
\rowcolor[gray]{.9} %
Tangent%
&All%
&Wikipedia (old subset)%
&Presentation%
&Custom%
&\footnotesize{http://saskatoon.cs.rit.edu/tangent/}\\ %
Uniquation %
&Math%
&Forums,%
&Presentation%
&Unknown%
&\footnotesize{http://uniquation.com/en/}\\ %
& %
&collaborative encyclopedias%
& %
& %
& \\%
\rowcolor[gray]{.9} %
Wolfram Alpha %
&Math-Sci%
&Encyclopedia, curated.%
&Meaning%
&\emph{Mathematica}%
&\footnotesize{https://www.wolframalpha.com/}\\ %
\rowcolor[gray]{.9} %
&Math%
&(Also generates results.)%
& %
& %
& \\ %
\bottomrule %
\end{tabular}
\begin{tablenotes}
      \footnotesize
      \item[A] ``Math-Sci'' indicates ``Mathematical Sciences''; for each engine listed here, that includes physics content.
    \end{tablenotes}
\end{threeparttable}
\end{table}
\begin{adjustwidth}{0in}{0in}
\begin{minipage}{0.79\linewidth}
\begin{table}[H]
\caption{Querying in Web-Available Engines}
\label{table:web-available-query}
\small
\begin{tabular}{SlSlSlSlSlSlSl}  \toprule
Search Engine & Language & \$$^1$ & $>$1 eqn.$^2$ & Keywords$^3$ & Loc.$^4$ & Wildcards (un-named$^5$) \\ \midrule 
\rowcolor[gray]{.9} %
LaTeXSearch%
&LaTeX%
&No%
&No%
&No%
&Yes%
&No$^6$\\ %
zbMATH%
&LaTeX%
&No%
&No%
&Yes%
&No%
&Preface symbol with ``?''.\\  %
&MathML%
&%
& %
& %
& %
&(Named wildcard.)\\%
\rowcolor[gray]{.9} %
EuDML%
&TeX%
&\$ or%
&No%
&Yes%
&Yes%
&No$^7$\\ %
\rowcolor[gray]{.9} %
&MathML%
&\$\$%
& %
& %
&%
& \\ %
DLMF%
&LaTeX-like%
&No%
&Yes%
&Yes%
&N/A
&``\$'' for many characters,\\ %
&%
&%
&%
& %
&%
& ``?'' for 0 or 1 characters.\\ %
\rowcolor[gray]{.9} %
Tangent%
&TeX%
&No%
&Yes%
&No%
&No%
&``?'' for single symbol.\\ %
Uniquation %
&TeX%
&No%
&No%
&No%
&No%
&Natively$^7$\\ %
\rowcolor[gray]{.9} %
Wolfram Alpha %
&Natural Language,%
&No%
&No$^8$%
&Yes%
&No%
&No$^6$\\%
\rowcolor[gray]{.9} %
&\emph{Mathematica},%
&%
& %
& %
&%
& \\ %
\rowcolor[gray]{.9} %
&LaTeX%
&%
&%
& %
& %
& \\ %
\bottomrule %
\end{tabular}
\end{table}
\end{minipage}%
\begin{minipage}{0.21\linewidth}
\footnotesize
\vspace{0.4em}
$^1$ Bracket text with \$'s, as for\\
\phantom{$^1$ }TeX math mode?\vspace{0.05em}\\
$^2$ Query can have $>$1 equation.\vspace{0.05em}\\
$^3$ Query can have keywords\\
\phantom{$^3$ }and equation(s).\vspace{0.05em}\\
$^4$ Specify location (e.g. in title)\\
\phantom{$^4$ }of query terms.\vspace{0.05em}\\
$^5$ Unless specified.\vspace{0.05em}\\
$^6$ Not specified; experiment-\\
\phantom{$^6$ }ation excluded ``\$'' and ``?''.\vspace{0.05em}\\
$^7$ Ignores variable names.\vspace{0.05em}\\
$^8$ Information not available;\\
\phantom{$^7$ }experimentation indicates\\
\phantom{$^7$ }``no''.  This applies to\\
\phantom{$^7$ }encylopediac information\\
\phantom{$^7$ }searches using equations,\\
\phantom{$^7$ }and not to solving equations\\
\phantom{$^7$ }directly with this tool.\\
\end{minipage}

\end{adjustwidth}

\newpage
\begin{table}[H]
\begin{threeparttable}
\caption{Downloadable Versions (\& one Physics-Catered Upcoming Engine)}
\small
\centering
\label{table:downloadable-main}
\begin{tabular}{SlSlSlSlSlSlSlSl}  \toprule %
Search Engine & Application & Doc. Encoding\tnote{A} & Crawler/Local & Search type & Web-server instructions & Download & Demo \\ \midrule 
\rowcolor[gray]{.9} %
Tangent%
&Custom database%
&TeX%
&Local%
&Presentation%
&Yes%
&\cite{dprl-descriptions,Tangent-github}%
&\cite{TangentSE}\\ %
\rowcolor[gray]{.9}%
&%
&%
&MathML%
&%
&%
&%
&\\ %
MathWebSearch%
&Custom database,%
&MathML%
&Web,%
&Meaning%
&Interface-dependant%
&\cite{MWSgithub}%
&\cite{zbMATH-engine,Sentido-search}\\ %
&Ready for arXiv%
&%
&MathML%
&%
&Local%
&%
&\\ %
\rowcolor[gray]{.9} %
MIaS%
&Custom database%
&MathML%
&Local%
&Presentation\tnote{B}%
&Yes (JVM instructions)%
&\cite{MIaS-github,WebMIaS-github}%
&\cite{MIaS-Demo}\\ %
$5e^{x+y}$%
&In-development,%
&LaTeX%
&Unknown%
&Meaning%
&N/A%
&N/A%
&N/A\\ %
\bottomrule %
\end{tabular}
\begin{tablenotes}
      \footnotesize
      \item[A] Documents whose math is encoded using these languages can be indexed for math-aware search.
      \item[B] Work is proposed to investigate including \emph{Mathematica}-like software in MIaS's meaning-recognition algorithms.
    \end{tablenotes}
\end{threeparttable}
\end{table}
\begin{table}[H]
\begin{threeparttable}
\caption{Querying in Downloadable Versions}
\label{table:downloadable-querying}
\small
\centering
\begin{tabular}{SlSlSlSlSlSl}  \toprule
Search Engine & Language & \$\tnote{1} & $>$1 eqn.\tnote{2} & Keywords\tnote{3} & Wildcards\tnote{4} \\ \midrule 
\rowcolor[gray]{.9} %
Tangent%
&TeX%
&No%
&Yes%
&Yes%
&``?''  for a symbol\\ %
MathWebSearch%
&Varies\tnote{5}%
&Varies\tnote{5}%
&No%
&Yes%
&Named: ``?'' before a symbol\\ %
\rowcolor[gray]{.9} %
MIaS%
&LaTeX%
&Yes, \$%
&Yes%
&Yes%
&``?''  for a symbol\\ %
\rowcolor[gray]{.9} %
&MathML%
&%
&%
&%
&\\ %
\bottomrule %
\end{tabular}
\begin{tablenotes}
\footnotesize
	\item[1] Bracket text with \$'s or \$\$'s, as for TeX math mode?
	\item[2] Query can have $>$1 equation.
	\item[3] Query can have keywords and equation(s).
	\item[4] Wildcards are un-named unless specified.
	\item[5] Depends on the interface which this engine is paired with.
\end{tablenotes}
\end{threeparttable}
\end{table}

\end{landscape}

\chapter{Examples}
\label{chap:screenshots}

A range of simple and neat equation queries are shown in this chapter, including at least one from each of the working web-based search engines.  A few features of the search interfaces are also pointed out.

\section{$m_{in}$}

\begin{figure}[H]
	\centering
	\makebox[\linewidth]{\includegraphics[width=11cm,frame]{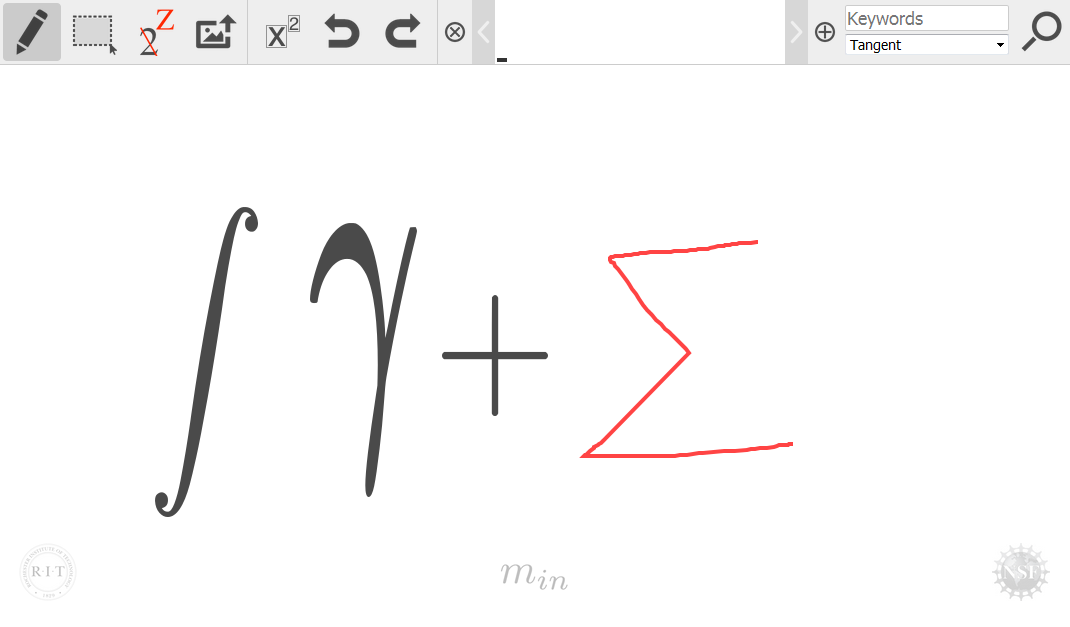}}
	\caption{Drawing in-progress, in $m_{in}$.}
	\label{fig:min-screen}
\end{figure}

The query in figure \ref{fig:min-screen} is in the process of being drawn, using $m_{in}$.  Instructions for $m_{in}$'s use are located at \cite{Min-instructions}.

\newpage
\section{LaTeXSearch}

\begin{figure}[H]
	\centering
	\makebox[\linewidth]{\includegraphics[width=15.5cm,frame]{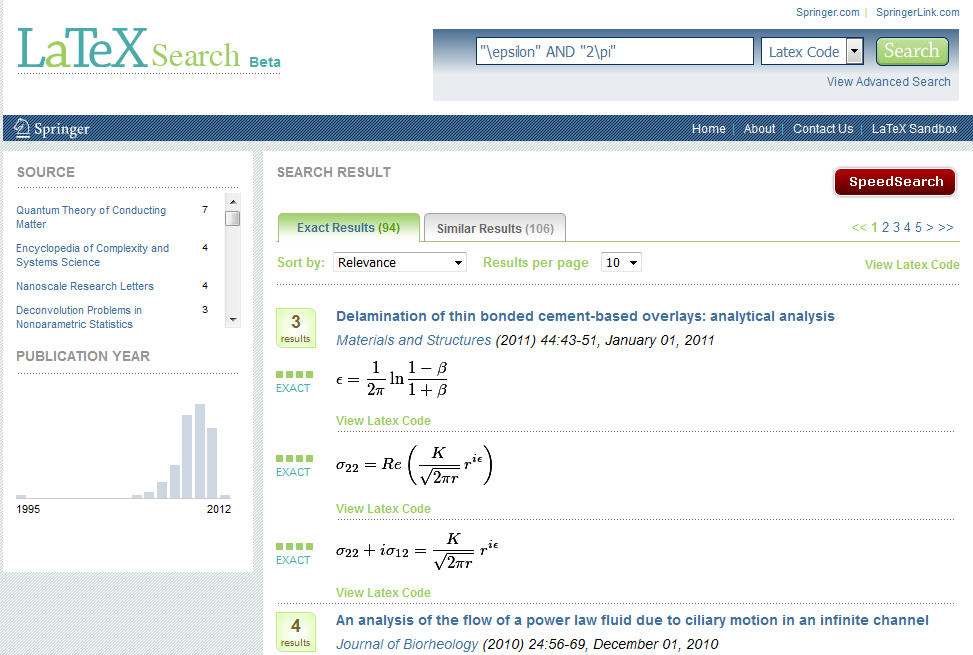}}
	\caption{A simple search on LaTeXSearch: ``\textbackslash{}epsilon'' AND ``2\textbackslash{}pi''.}
	\label{fig:LatexSearch-screen}
\end{figure}

In the simple search shown in figure \ref{fig:LatexSearch-screen}, note the post-search filters located in three locations: within the tabs, just below the tabs, and in the left vertical side-bar.  In the display, equations from each paper are grouped together; the number of equations in each paper is listed to the left of the publication's name.  LaTeX code can be downloaded via the green ``View Latex Code'' links.

\newpage
\section{zbMATH}

\begin{figure}[H]
	\centering
	\makebox[\linewidth]{\includegraphics[width=14cm,frame]{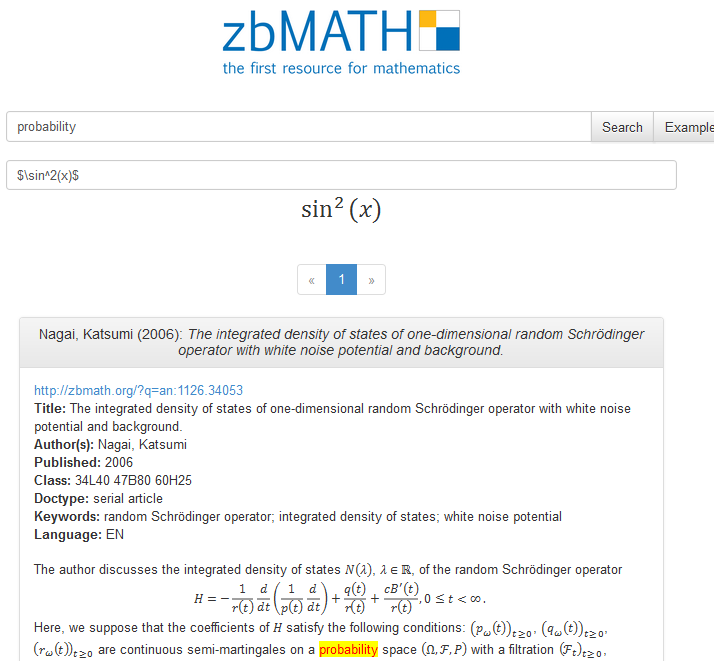}}
	\caption{A keyword (``probability'') and equation ($\sin^2(x)$) search on the zbMATH platform.  The formula searched for is also present in the results' display, but it does not fit into this screen capture.}
	\label{fig:zbMath-screen}
\end{figure}

A neat combined keyword and equation search in zbMATH is shown in figure \ref{fig:zbMath-screen}.

When generating this screenshot, it was necessary to click on the result's grey title box to show its full entry.  The equation matched, ($\sin^2(x)$) does appear in the results, but is too low down to include in the screenshot.

\newpage
\section{EuDML}

\begin{figure}[H]
	\centering
	\makebox[\linewidth]{\includegraphics[width=14cm,frame]{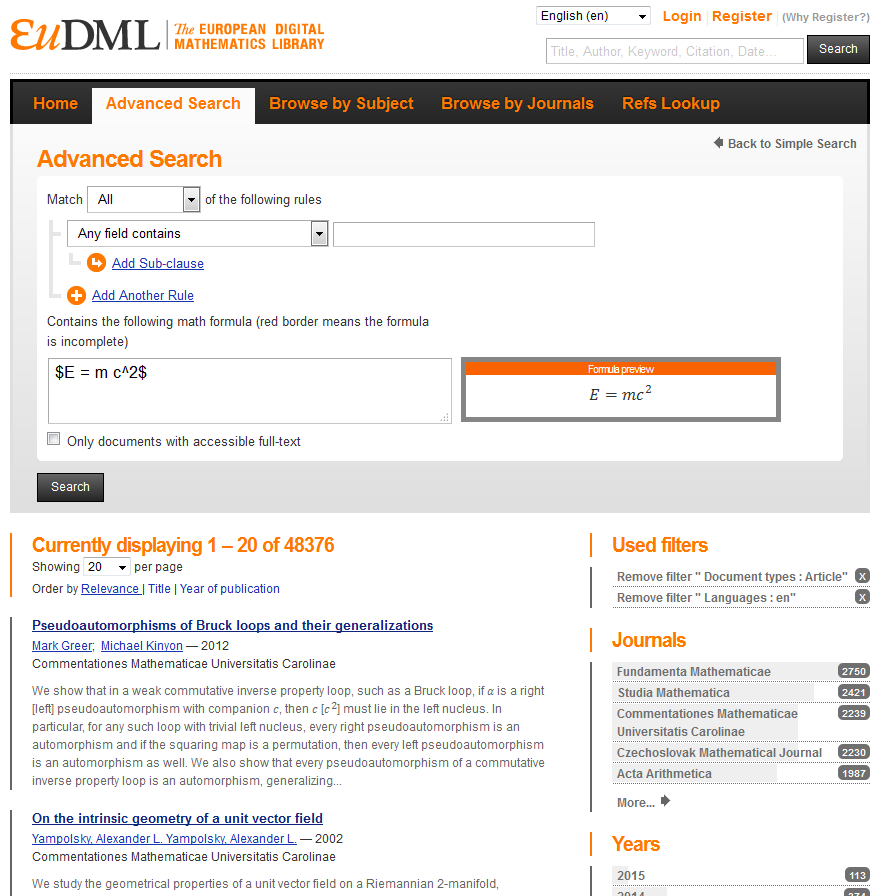}}
	\caption{A sample search in EuDML: $E = m c^2$.  Other interesting result titles, lower on the page, included ``The equivalence principle meets the uncertainty principle'' (1988), and ``On the validity of Wilson's approach to general relativity'' (1997), and ``Recovering Asymptotics at Infinity of Perturbations of Stratified Media'' (2000).  Whether these are exact or similar formula matches is unknown: given the search setup, determining this requires accessing and downloading the article.}
	\label{fig:euDML-screen}
\end{figure}

Using the advanced search function for EuDML, the equation $E = m c^2$ was queried.  Putting post-search filters of ``English'' and ``Articles'' produced the result in Figure \ref{fig:euDML-screen}.

\newpage
\section{DLMF}

The queries used here may not work in other search engines, since they were drawn from the DLMF help webpage. \cite{DLMF-help}

\begin{figure}[H]
	\centering
	\begin{subfigure}[a]{\textwidth}
		\makebox[\linewidth]{\includegraphics[width=14cm,frame]{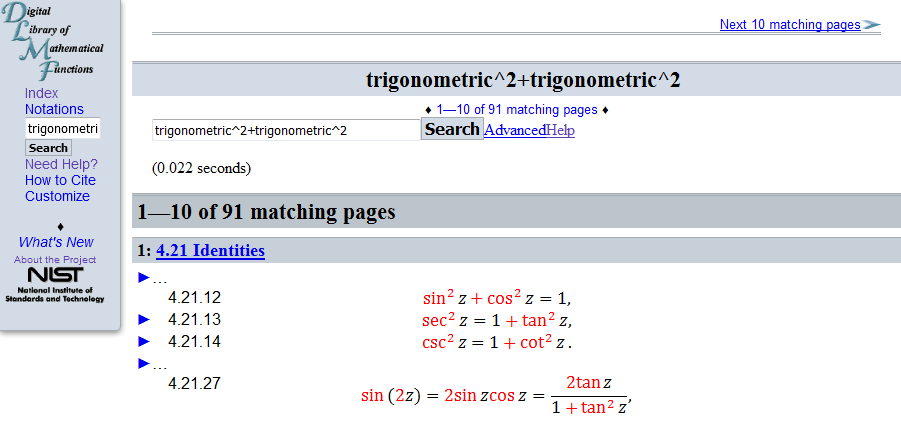}}
	\end{subfigure}
	\begin{subfigure}[b]{\textwidth}
		\makebox[\linewidth]{\includegraphics[width=14cm,frame]{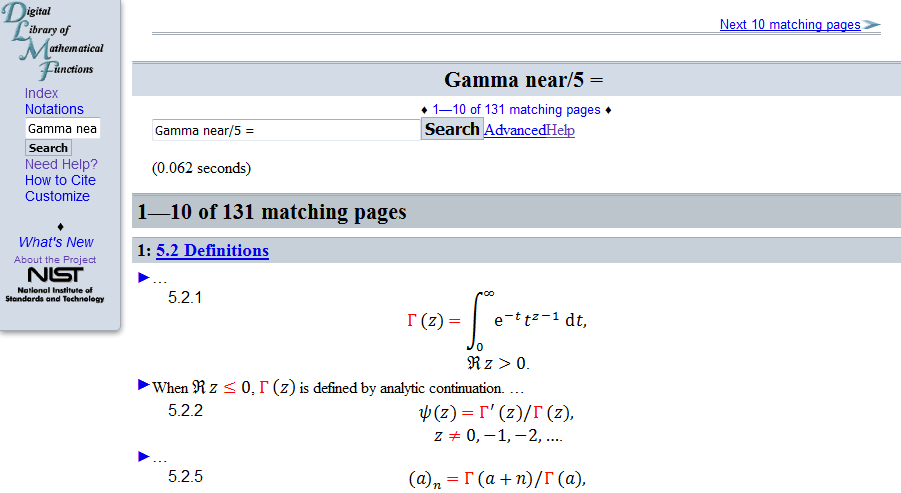}}
	\end{subfigure}
	\caption{Tricky search examples, using language which is specific to the DLMF.  The top panel query is ``trigonometric\textasciicircum{}2 + trigonometric\textasciicircum{}2''; the bottom is ``Gamma near/5 =''.}
	\label{fig:DLMF-screen}
\end{figure}

In figure \ref{fig:DLMF-screen}'s upper panel, the search is for the sum of the squares of any two trigonometric functions.  Its lower panel shows a search for equations with $\Gamma$ and $=$ both present and separated by no more than 5 terms.

\newpage
\section{Tangent}

\begin{figure}[H]
	\makebox[\linewidth][c]{%
	\begin{subfigure}{15.5cm}
		\begin{subfigure}{7.75cm}
			\makebox[\linewidth]{\includegraphics[width=7.75cm,frame]{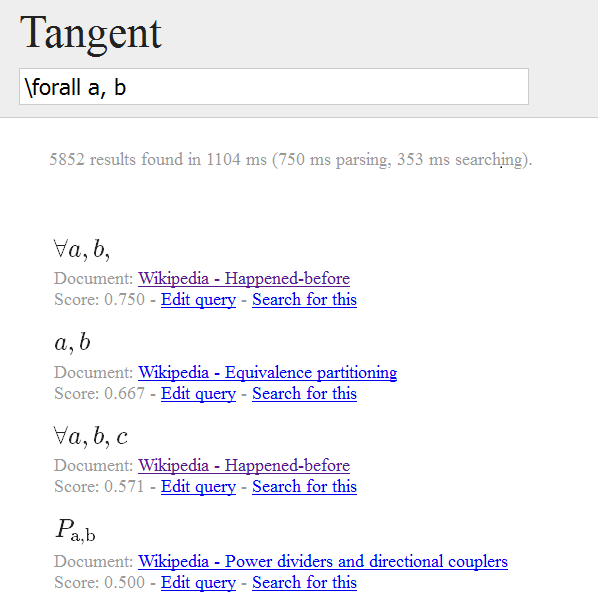}}
		\end{subfigure}
		\begin{subfigure}{7.75cm}
			\makebox[\linewidth]{\includegraphics[width=7.75cm,frame]{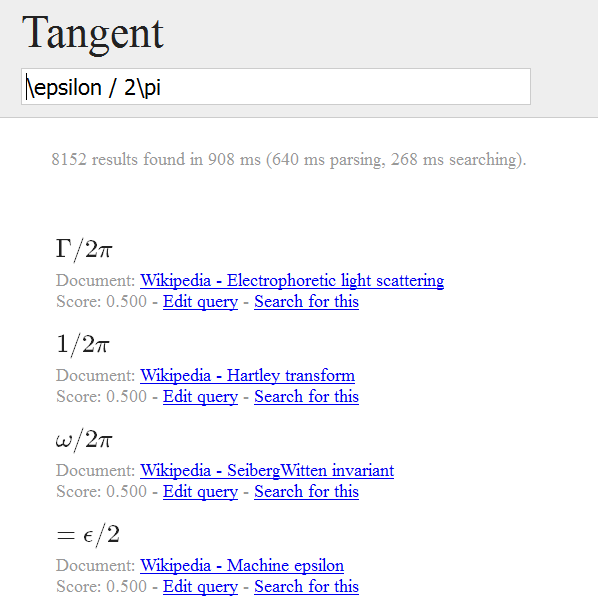}}
		\end{subfigure}
	\end{subfigure}
	}\\
	\caption{Neat and simple search query examples, and a showcase of the Tangent platform.  The left panel's query is "\textbackslash{}forall a,b", and the right is "\textbackslash{}epsilon / 2\textbackslash{}pi".}
	\label{fig:Tangent-screen}
\end{figure}

Simple queries are shown in figure \ref{fig:Tangent-screen}, on the Tangent search engine platform.

\newpage
\section{Uniquation}
\label{sec:uniquation-screenshots}

\begin{figure}[H]
	\centering
	\begin{subfigure}[a]{\textwidth}
		\makebox[\linewidth]{\includegraphics[width=12cm,frame]{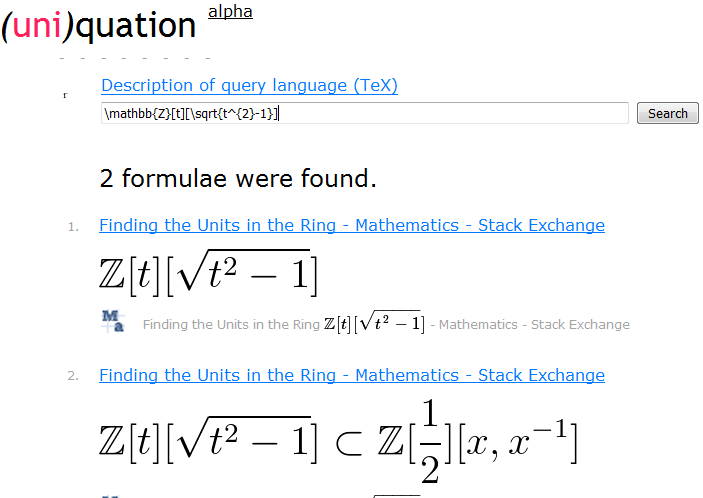}}
	\end{subfigure}
	\begin{subfigure}[b]{\textwidth}
		\makebox[\linewidth]{\includegraphics[width=12cm,frame]{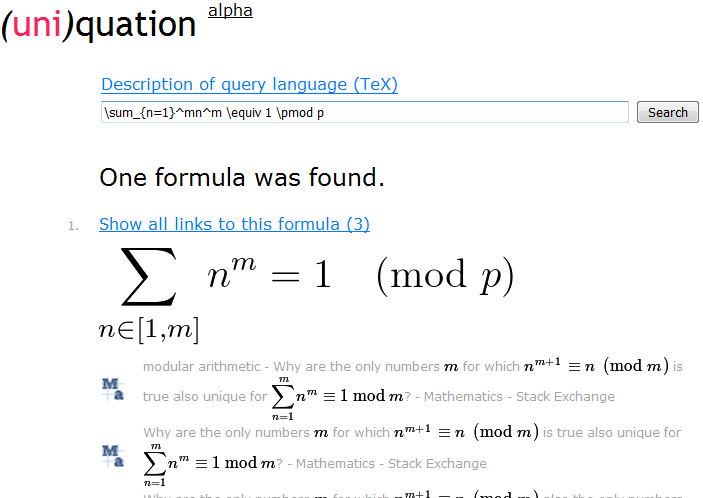}}
	\end{subfigure}
	\caption{Examples of neat searches, using Uniquation.  The query in the top panel is ``\textbackslash{}mathbb\{Z\}[t][\textbackslash{}sqrt\{t\textasciicircum{}2 - 1\}]'', and that in the bottom panel is ``\textbackslash{}sum\_\{n=1\}\textasciicircum{}m n\textasciicircum{}m \textbackslash{}equiv 1 \textbackslash{}pmod p''. Examples are drawn from \cite{uniquation-twitter}.}
	\label{fig:Uniquation-screen}
\end{figure}

Uniquation's inclusion of abstract algebra (top panel) and a general example of how sums might be searched for (bottom panel) are both shown in figure \ref{fig:Uniquation-screen}.  Note that results are grouped by the formula which they contain; this is evident in the lower panel of figure \ref{fig:Uniquation-screen}, where two results are shown under formula ``1''.

\newpage
\section{Wolfram Alpha}

\begin{figure}[H]
	\centering
	\makebox[\linewidth]{\includegraphics[width=14cm,frame]{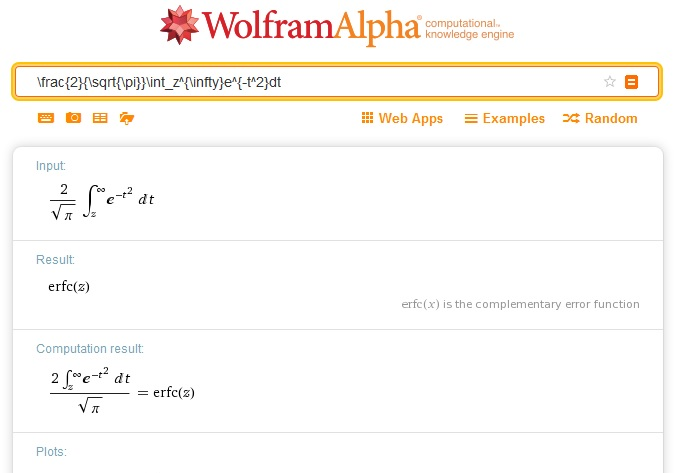}}
	\caption{A Wolfram Alpha search, using a LaTeX-encoded query, for the integral representation of the $\erfc(z)$.}
	\label{fig:WolframAlpha-screen}
\end{figure}

A search for the complementary error function's integral representation,
\begin{equation}
	\frac{2}{\sqrt{\pi}}\int_z^{\infty}e^{-t^2}dt
\end{equation}
is shown in figure \ref{fig:WolframAlpha-screen}.  Wolfram Alpha correctly identifies this as $\erfc(z)$.

\chapter{Conclusion}

New research methods are enabled by equation-based search engines.  From the simple benefit of looking up a concept by its equation - whether half-remembered or fully - to the advanced application of helping researchers discover new interdisciplinary connections, the possible applications elucidated so far have likely barely scratched the surface.

Meaning-based equation searches can open new doors to math solutions, perspectives, collaborations, and faster-running computer code.  Presentation-based searches can help with the same,except that they require equations' structures to be similar in order for a match to be found; they can also help with finding out what unfamiliar math means.

These uses of math-aware search engines can lead to better understanding, predictability, and applications of physical systems.  Fields which may see progress accelerated by these engines range from fundamental physics (e.g. gravitation and quantum physics) to effective theories (e.g. within condensed matter and fluid mechanics) to technological applications (e.g. quantum computing and energy extraction).

Databases which are presently searchable in this way include journal articles in physics, math, and other mathematical disciplines (LaTeXSearch, EuDML, zbMATH); Wikipedia (Tangent); a curated math encyclopedia which includes some physics (Wolfram Alpha); and math forums (Uniquation).  Expansion to the entire arXiv.org website is feasible with MathWebSearch, a meaning-based math-aware search engine.  Other expansions are also planned.  A physics-specific engine is also in development, at CERN, which would search CERN's document server using math-aware equations; it takes into account some specific desires of physicists, during the math matching stage.

Other developments planned include improving the: ranking algorithms for results, mathematical-meaning extraction and comparison, and speed of retrieval (which is passable but not as fast as Google).  A different sort of search which is also being considered is search by diagram.

The search tools and discussion in this document provide a resource for physicists (and other researchers) to encompass the language of math in their searches.  Whether these math-aware search engines are used for targetted information searches, exploration, or other inspiration, they are sure to prove useful in physics.

\appendix
\chapter{Glossary}
\label{app:glossary}
\settocdepth{chapter}

Terms used in this discussion of math-aware search engines are defined here.  Many of these terms are not yet set into conventions, in this field.  Some of the definitions may seem very straightforward, but can have varying definitions between and within research fields.

\paragraph{Math-Aware Search Engine / Math-language search engine}
A software which searches a database of mathematical texts, using a mathematical equation as at least part of its search query.  It accounts, at least partially, for the mathematical equivalencies between different writings of equations.

\paragraph{Query}
The set of items which the user inputs as search terms.  It can consist, for example, of equations, keywords, phrases, DOIs, and publication date range.

\paragraph{Database}
A collection of data.  For equation-based search engines, it might consist of journal articles, wikis, forums, encyclopedias, or other documents.

\paragraph{Equation match}
An equation which matches another equation, within the search engine’s criteria.  For example, the following equations might match: $a x+5 = b$, and $s x + 5 = t$.

\paragraph{Result}
The items from the database (e.g. equations, pieces of text) which match the search query given the algorithms of the search engine.

\paragraph{Rank of a result}
The relevance of a database equation, relative to the query.  It is described by a number. Results with a higher rank are located higher in the list of results. \cite{Guidi-Cohen-2015, Tangent-pub-2015, SojkaEtAl2011,  Rec-and-retrieval-rev}

\paragraph{Subformula}
Pieces of a formula.  For example, $a+b^{2+c}$ has two subformulae: $b^{2+c}$ and $2+c$.  The expression $a+b^{2+c}$ is itself a subformula of $2 / (a+b^{2+c})$. \cite{Oviedo-Th, SojkaEtAl2011}

\paragraph{Meaning-based search}
Searches based on the true mathematical meaning of an equation, with the goal of identifying all equations which have equivalent or similar meanings to the query equation.  For example, the meaning of $\sinh(x)$ and $(1/2)(e^z-e^{-z})$ are equal.  \cite{Guidi-Cohen-2015}  Meaning-based search engines which are now web-ready obtain only part of this goal, but they do account for some mathematical meaning. \cite{Pres-vs-Content-2014, MWS-05-2012, AizawaEtAl2014}

\paragraph{Presentation-based search}
Bases search results on the visual appearance of an equation.  For example, $1 + x^2$ is visually different from $x^2 + 1$.  The equations $1+ x^2 /2$ and $1+ y^2 /2$ match visually, so would count as similar equations in a presentation-based search.  Some presentation-based search engines account for algebraic rearrangement of terms (e.g., $1 + x^2$ versus $x^2 + 1$), and others do not. \cite{Tangent-technical-v0.3, TangentSE, Liska-Th, Pres-vs-Content-2014}

\chapter{Nifty Math Tools}
\label{app:math-tools}

There is a lot of scientific software which is publicly available, but it can be hard to find unless a researcher knows exactly what to look for.

This appendix covers a sample of interesting math tools, and gives references to a few good databases and lists of tools.  These were discovered along the course of this research, and the author thought them nifty: the list is by no means exhaustive.

\section{Math tools: A Sample of Interesting Tools}

Below is a subset of Mathematical software tools: it includes a few which I thought might be of interest to physicists, but is intended to be a sample only.  In the next section (section \ref{sec:finding-tools}), some resources for finding other math software are given.

\subsection{Number Knowledge Lookup}

\subsubsection{Inverse Symbolic Calculator}
\label{sec:inverse-symbolic-calc}
Enter a decimal number.  The tool tries to find combinations of mathematical constants which could make up that number.  It ignores factors of 10; why this is so is not stated on the webpage. \footnote{It is not clear to me why having 10 of something should be special, as opposed to having, say 8 of it: why are factors of 10 more important than are factors of 8, or 4, or 3?  However, perhaps the factor of 10 is special, given our base-10 number system, from number theory.} \cite{inverse-symb-calc}
 
\subsubsection{The On-Line Encyclopedia of Integer Sequences}
Enter the first few numbers of an integer sequence; this tool will try to identify whether it is part of a formal sequence, and bring up references for it.  It is recommended, on the ``hints'' page, that about 6 terms in the sequence be used, starting with the second or third term. \cite{integer-seq-encyc}

\subsubsection{Encyclopedia of Combinatorial Structures}
This site is like the Online Encyclopedia of Integer Sequences, except that it focuses on sequences which are relevant to decomposable combinatorial structures.  Search options include: the first few terms from a sequence, keyword, generating function, or closed form. \cite{combinatorial-encyc}

\subsection{Exploration and/or Lookup}

\subsubsection{The KnotPlot Site}
Contains a collection of knots and links images, and a link to a software by which to visualize and manipulate three-and four-dimensional mathematical knots. \cite{knot-plot}

\subsubsection{Interactive Geometry Software: Cinderella}
Designed for university-level math, this free downloadable software helps users explore geometry visually. \cite{Cinderella}

\subsection{Proof Assistants}
Mathematical proof assistants include the semi-interactive theorem prover Coq \cite{Coqweb} and the coding language Mizar \cite{Mizar-web}. \cite{Guidi-Cohen-2015}

\subsection{Special and Elementary Function databases}
Aside from the equation-searchable special function databases, which were mentioned in the main paper (DLMF, Wolfram Alpha), there are some keyword-searchable special function databases.  One example is the Wolfram Functions website, where a special function's entry can contain many different definitions and written forms.  A keyword by which to search for more such sites is ``Digital Mathematics Library'' (DML).

There is also an in-development upgrade to searching special functions and orthogonal polynomials, which includes the content of the DLMF and a number of special functions textbooks.  It is planned to be easily searchable \cite{DRMF-CohlEtAl}; since it aims to improve upon the DLMF, hopefully this will include a math-aware search engine which can parse equation queries.  Community comments on each formula's page will also be possible.  A demo of it is available at \cite{DRMF-web-Xsede} or \cite{DRMF-web-wmflabs}.

\subsection{Modeling tools}
Many modeling tools are listed on swMath.org.  A few examples are the modeling-oriented coding language Modelica (thank you to Erik Schnetter for pointing out this tool) \cite{modelica}, and Wolfram's Systems Modeler  platform \cite{Wolfram-System-Modeler}.

\section{Resources to find other tools}
\label{sec:finding-tools}

\subsection{Database of Custom Scientific \& Mathematical Software}
swMath.org \cite{swMATH-web} collects subject-browsable software which has been made for researchers.  It can also be searched by keyword; software name, authors, and description; programming language; and category of research (``MSC Classification'').  Covered subjects include:
\begin{itemize}
\item	Quantum theory
\item	Astronomy and astrophysics
\item	Approximation and expansions
\item	Numerical analysis
\item	Other math and physics subjects
\end{itemize}
References are included to publications which discuss or use each software; the number of such ``citations'' of a software is presented in search results.

Some of the information in swMath.org is outdated, in that the software is no longer available and/or operational.  But, it is a great place to start for researchers who seek knowledge about whether software solutions exist for their particular problem.  For general browsing, it is also very informative.

\subsection{Lists of Math Tools}

The following places have good lists of references to other tools, and/or to other lists.  They are ones which I happened upon; many other good lists undoubtedly exist.
\begin{itemize}
\item	\cite{Newcastle-Jon-ref}: a resource list by the developer of the Inverse Symbolic Calculator (see section \ref{sec:inverse-symbolic-calc}).
\item	\cite{devel-global-DML}: the section ``Specialized Mathematical Databases'', and the subsequent couple of sections, starting at page 120.
\end{itemize}

\footnotesize{\bibliography{bibtex_MSc_proj_Apr20}}

\end{document}